\documentclass[prb]{revtex4}

\usepackage{graphicx}
\usepackage{dcolumn}
\usepackage{bm}
\usepackage{subfigure}
\usepackage{amssymb}
\usepackage{graphics}
\usepackage{epsfig}
\usepackage{threeparttable}
\newcommand{\thickhline}{\noalign{\hrule height 0.8pt}}

\usepackage{threeparttable}

\begin{document}

\preprint{APS/123-QED}

\title{Critical behavior and phase diagrams of a spin-1 Blume-Capel model with random crystal field interactions: An effective field theory analysis  }
\author{Yusuf Y\"{u}ksel}%
\altaffiliation[Also at ]{Dokuz Eyl\"{u}l University,
Graduate School of Natural and Applied Sciences, Turkey}
\author{\"{U}mit Ak{\i}nc{\i}}
\affiliation{Department of Physics, Dokuz Eyl\"{u}l University,
TR-35160 Izmir, Turkey}
\author{Hamza Polat}
\email{hamza.polat@deu.edu.tr}
\affiliation{Department of Physics, Dokuz Eyl\"{u}l University,
TR-35160 Izmir, Turkey}


\date{\today}

\begin{abstract}
A spin-1 Blume-Capel model with dilute and random crystal fields is examined for honeycomb and square lattices by introducing an effective-field approximation that takes into account the correlations between different spins that emerge when expanding the identities. For dilute crystal fields, we have given a detailed exploration of the global phase diagrams of the system in $k_{B}T_{c}/J-D/J$ plane with the second and first order transitions, as well as tricritical points. We have also investigated the effect of the random crystal field distribution characterized by two crystal field parameters $D/J$ and $\triangle/J$ on the phase diagrams of the system. The system exhibits clear distinctions in qualitative manner with coordination number $q$ for random crystal fields with $\triangle/J,D/J\neq0$. We have also found that, under certain conditions, the system may exhibit a number of interesting and unusual phenomena, such as reentrant behavior of first and second order, as well as a double reentrance with three successive phase transitions.
\begin{description}
\item[\qquad \qquad \qquad PACS numbers]
75.10.Dg, 75.10.Hk, 75.30.Kz
\end{description}
\end{abstract}

\maketitle
\tableofcontents

\section{Introduction}\label{introduction}
Spin-1 Blume-Capel (BC) model \cite{blume,capel} is one of the most extensively studied models in statistical mechanics and condensed matter physics. The model exhibits a variety of multicritical phenomena such as a phase diagram with ordered ferromagnetic and disordered paramagnetic phases separated by a transition line that changes from a continuous phase transition to a first-order transition at a tricritical point. On the other hand, as an extension of the model, BC model with a random crystal field represents the critical behavior of $\mathrm{^{3}He}-\mathrm{^{4}He}$ mixtures in a random media, i.e., aerogel where $S=0$ and $S=\pm1$  states represent $\mathrm{^{3}He}$ and $\mathrm{^{4}He}$ atoms, respectively \cite{aerogel1,aerogel2}. From the theoretical point of view, BC model with a random crystal field (RCF) has been studied by a variety of techniques such as cluster variational method (CVM) \cite{aerogel2}, Bethe lattice approximation (BLA) \cite{albayrak}, effective field theory (EFT) \cite{kaneyoshi1,kaneyoshi2,kaneyoshi3,kaneyoshi4,yan}, finite cluster approximation (FCA) \cite{benyoussef1,ilkovic}, mean field theory (MFT) \cite{benyoussef2,borelli,carneiro2,boccara,carneiro1,bahmad}, Monte Carlo (MC) simulations \cite{puha}, pair approximation (PA) \cite{lara}, and renormalization group (RG) method \cite{branco}. Among these studies, EFT and MFT have been widely used to investigate the thermal and magnetic properties of BC model with a RCF distribution. For example, Kaneyoshi and Mielnicki \cite{kaneyoshi3} investigated the phase diagram of the system for a honeycomb lattice by using EFT with correlations and they found some important differences from the results obtained by the standard MFT. Similarly, in a recent paper, Yan and Deng \cite{yan} considered the same model within the framework of EFT, and they derived the expressions of magnetizations for honeycomb and square lattices. On the other hand, in several studies based on MFT, \cite{benyoussef2,carneiro2} the authors paid attention to the effects of crystal field dilution on the phase diagrams of the system and they observed that the system may exhibit a reentrant behavior, as well as first order phase transitions.

However, EFT and MFT studies mentioned above have some unsatisfactory results. Namely, the results obtained by EFT are limited to second-order phase transitions and tricritical points, and a detailed description of first-order transitions has not been reported. Other than this, it is well known that magnetic systems with dilute crystal fields exhibit qualitatively similar characteristics when compared to site dilution problem of magnetic atoms. From this point of view, for a BC model with diluted crystal fields, MFT predicts that the phase transition temperature of the system will remain at a finite value until zero concentration is reached. In this context, we believe that BC model with RCF still deserves particular attention for investigating the proper phase diagrams, especially the first-order transition lines that include reentrant phase transition regions. Conventional EFT approximations include spin-spin correlations resulting from the usage of the Van der Waerden identities, and provide results that are superior to those obtained within the traditional MFT. However, these conventional EFT approximations are not sufficient enough to improve the results, due to the usage of a decoupling approximation (DA) that neglects the correlations between different spins that emerge when expanding the identities. Therefore, taking these correlations into consideration will improve the results of conventional EFT approximations. In order to overcome this point, recently we proposed an approximation that takes into account the correlations between different spins in the cluster of a considered lattice \cite{akýncý}. Namely, an advantage of the approximation method proposed by this study is that no decoupling procedure is used for the higher-order correlation functions.

In this paper, we intent to investigate the effects of RCF distributions on the phase diagrams of spin-1 BC model on 2D lattices, namely honeycomb $(q=3)$ and square $(q=4)$ lattices. For this purpose, the paper is organized as follows: In Sec. \ref{formulation} we briefly present the formulations. The results and discussions are presented in Sec. \ref{results}, and finally Sec. \ref{conclude} contains our conclusions.

\section{Formulation}\label{formulation}
In this section, we give the formulation of the present study for a 2D lattice which has $N$ identical spins arranged. We define a cluster on the lattice which consists of a central spin labeled $S_{0}$, and $q$ perimeter spins being the nearest neighbors of the central spin. The cluster consists of $(q + 1)$ spins being independent from the spin operator $\hat{S}$. The nearest-neighbor spins are in an effective field produced by the outer spins, which can be determined by the condition that the thermal average of the central spin is equal to that of its nearest-neighbor spins. The Hamiltonian describing our model is
\begin{equation}\label{eq1}
H=-J\sum_{\langle i,j\rangle}S_{i}^{z}S_{j}^{z}-\sum_{i}D_{i}(S_{i}^{z})^{2},
\end{equation}
where the first term is a summation over the nearest-neighbor spins with $S_{i}^{z}=\pm1, 0$ and the term $D_{i}$ on the second summation represents a random crystal field, distributed according to a given probability distribution. In this paper, we primarily deal with two kinds of probability distributions, namely, a quenched diluted crystal field distribution and a double peaked delta distribution which are given by Eqs. (\ref{eq2}) and (\ref{eq3}), respectively as follows
\begin{eqnarray}\label{eq2}
P(D_{i})&=&p\delta(D_{i}-D)+(1-p)\delta(D_{i}),\\
\label{eq3}
P(D_{i})&=&\frac{1}{2}\left\{\delta[D_{i}-(D-\triangle)]+\delta[D_{i}-(D+\triangle)]\right\}.
\end{eqnarray}
where $p$ denotes the concentration of the spins on the lattice which are influenced by a crystal field $D$.

We can construct the mathematical background of our model by using the approximated spin correlation identities \cite{sa_barreto_et_al}
by taking into account random configurational averages
\begin{equation}\label{eq4}
\langle\langle
\{f_{i}\}S_{i}^{z}\rangle\rangle_{r}=\left\langle\left\langle
\{f_{i}\} \frac{\mathrm{Tr}_{i}\left(S_{i}^{z}\right)\exp{(-\beta
H_{i})}}{\mathrm{Tr}_{i}\exp{(-\beta H_{i})}}\right\rangle\right\rangle_{r},
\end{equation}
\begin{equation}\label{eq5}
\langle\langle
\{f_{i}\}(S_{i}^{z})^2\rangle\rangle_{r}=\left\langle\left\langle
\{f_{i}\} \frac{\mathrm{Tr}_{i}(S_{i}^{z})^2\exp{(-\beta
H_{i})}}{\mathrm{Tr}_{i}\exp{(-\beta H_{i})}}\right\rangle\right\rangle_{r},
\end{equation}
where $\beta=1/k_{B}T$, $\{f_{i}\}$ is an arbitrary function which is independent of the spin variable $S_i$ and the inner $\langle...\rangle$ and the outer $\langle...\rangle_{r}$ brackets represents the thermal and random configurational averages, respectively.

In order to apply the differential operator technique \cite{kaneyoshi_honmura,kaneyoshi5}, we should separate the Hamiltonian (\ref{eq1}) into two parts as $H=H_{i}+H^{'}$. Here, the effective Hamiltonian $H_{i}$ includes all the contributions associated with the site $i$, and the other part $H^{'}$ does not depend on the site $i$.
\begin{equation}\label{eq6}
-H_{i}=ES_{i}^{z}+D_{i}\left(S_{i}^{z}\right)^{2},
\end{equation}
where $E=J\sum_{j}S_{j}^z$ is the local field on the site $i$. If we use the matrix representations of the operators $S_{i}^{z}$ and $(S_{i}^{z})^{2}$ for the spin-1 system then we can obtain the matrix form of Eq. (\ref{eq6})
\begin{equation}\label{eq7}
-H_{i}=\left(
  \begin{array}{ccc}
    E+D & 0 & 0 \\
    0 & 0 & 0 \\
    0 & 0 & -E+D \\
  \end{array}
\right).
\end{equation}

Hereafter, we apply the differential operator technique in Eqs. (\ref{eq4}) and (\ref{eq5}) with $\{f_{i}\}=1$. From Eq. (\ref{eq4}) we obtain the following spin identity with thermal and configurational averages of a central spin for a lattice with a coordination number $q$ as
\begin{equation}\label{eq8} \langle\langle
S_{0}^{z}\rangle\rangle_{r}=\left\langle\left\langle\prod_{j=1}^{q}\left[1+S_{j}^{z}\mathrm{sinh}(J\nabla)+(S_{j}^{z})^{2}\{\mathrm{cosh}(J\nabla)-1\}\right]\right\rangle\right\rangle_{r}
F(x)|_{x=0}.
\end{equation}
The function $F(x)$ in Eq. (\ref{eq8}) is defined by
\begin{equation}\label{eq9}
F(x)=\int dD_{i}P(D_{i})f(x,D_{i}),
\end{equation}
where
\begin{eqnarray}\label{eq9a}
f(x,D_{i})&=&\frac{1}{\sum_{n=1}^{3}\exp(\beta\lambda_{n})}\sum_{n=1}^{3}\langle \varphi_{n}|S_{i}^{z}|\varphi_{n}\rangle\exp(\beta\lambda_{n}),\\
\nonumber
&=&\frac{2\sinh(\beta x)}{2\cosh(\beta x)+e^{-\beta D_{i}}}.
\end{eqnarray}
In Eq. (\ref{eq9a}), $\lambda_{n}$ denotes the eigenvalues of $-H_{i}$ matrix in Eq. (\ref{eq7}), and $\varphi_{n}$ represents the eigenvectors corresponding to the eigenvalues $\lambda_{n}$ of $-H_{i}$ matrix. With the help of Eq. (\ref{eq9a}), and by using the distribution functions defined in Eqs. (\ref{eq2}) and (\ref{eq3}), the function $F(x)$ in Eq. (\ref{eq9}) can be easily calculated by numerical integration. Hereafter, we will focus our attention on the construction of the correlation functions, as well as magnetization and quadrupole moment identities of a honeycomb lattice with $q=3$. A brief formulation of the fundamental spin identities for a square lattice with $q=4$ can be found in Appendix \ref{appendixa}.

By expanding the right-hand side of Eq. (\ref{eq8}) for a honeycomb lattice with $q=3$, we get the longitudinal magnetization as
\begin{eqnarray}\label{eq10}
\nonumber m_{z}=\langle\langle
S_{0}^{z}\rangle\rangle_{r}&=&l_{0}+3k_{1}\langle\langle S_{1}
\rangle\rangle_{r}+3(l_{1}-l_{0})\langle\langle S_{1}^{2}
\rangle\rangle_{r}+3l_{2}\langle\langle S_{1}S_{2}
\rangle\rangle_{r} \\
\nonumber& & +6(k_{2}-k_{1})\langle\langle S_{1}S_{2}^{2}
\rangle\rangle_{r}+3(l_{0}-2l_{1}+l_{3})\langle\langle
S_{1}^{2}S_{2}^{2}
\rangle\rangle_{r} \\
\nonumber& &+k_{3}\langle\langle S_{1}S_{2}S_{3}
\rangle\rangle_{r}+3(l_{4}-l_{2})\langle\langle S_{1}S_{2}S_{3}^{2}
\rangle\rangle_{r}\\
\nonumber& &+3(k_{1}-2k_{2}+k_{4})\langle\langle
S_{1}S_{2}^{2}S_{3}^{2}
\rangle\rangle_{r}\\
& &+(-l_{0}+3l_{1}-3l_{3}+l_{5})\langle\langle
S_{1}^{2}S_{2}^{2}S_{3}^{2} \rangle\rangle_{r},
\end{eqnarray}
We note that, for the sake of simplicity, the superscript $z$ is omitted from the correlation functions on the right-hand side of Eq. (\ref{eq10}). The coefficients in Eq. (\ref{eq10}) are defined as follows:
\begin{eqnarray}\label{eq11}
\nonumber  l_{0}=F(0), \ \ \ \ \ \ \ \ \ \ \ \ \ \ \ \ \ \ \ \ \ \ \ \ \ \ \ \ \ \ \ \ \ \ \ \ \ \ \ \ \ \  & &  \\
\nonumber  l_{1}=\mathrm{cosh}(J\nabla)F(x)|_{x=0},\ \ \ \ \ \ \ \ \ \ \ \ \ \ \ \ \ \ \ \ \ \ \ \ & & k_{1}= \mathrm{sinh}(J\nabla)F(x)|_{x=0},  \\
\nonumber  l_{2}=\mathrm{sinh}^{2}(J\nabla)F(x)|_{x=0},\ \ \ \ \ \ \ \ \ \ \ \ \ \ \ \ \ \ \ \ \ \ \ & & k_{2}= \mathrm{cosh}(J\nabla) \mathrm{sinh}(J\nabla)F(x)|_{x=0},  \\
\nonumber  l_{3}=\mathrm{cosh}^{2}(J\nabla)F(x)|_{x=0}, \ \ \ \ \ \ \ \ \ \ \ \ \ \ \ \ \ \ \ \ \ \ & & k_{3}= \mathrm{sinh}^{3}(J\nabla)F(x)|_{x=0}, \\
\nonumber  l_{4}=\mathrm{cosh}(J\nabla) \mathrm{sinh}^{2}(J\nabla)F(x)|_{x=0}, \ \ \ \ \ \ \ \ \ \ & & k_{4}=\mathrm{cosh}^{2}(J\nabla) \mathrm{sinh}(J\nabla)F(x)|_{x=0}, \\
l_{5}=\mathrm{cosh}^{3}(J\nabla)F(x)|_{x=0}, \ \ \ \
\ \ \ \ \ \ \ \ \ \ \ \ \ \ \ \ \ \ & &
\end{eqnarray}
Next, the average value of the perimeter spin in the system can be written as follows, and it is found as
\begin{equation}\label{eq12}
m_1= \langle\langle S_{\delta}^{z}\rangle\rangle_{r}=
\langle\langle 1+S_{0}^{z}\mathrm{sinh}(J\nabla)+(S_{0}^{z})^{2} \{\mathrm{cosh}(J\nabla)-1\}\rangle\rangle_{r}
F(x+\gamma)|_{x=0},
\end{equation}
\begin{equation}\label{eq13}
\langle\langle S_{1}\rangle\rangle_{r}=a_{1}\left(1-\langle\langle(S_{0}^{z})^{2}\rangle\rangle_{r}\right)+a_{2}\langle\langle S_{0}^{z}\rangle\rangle_{r}+a_{3}\langle\langle(S_{0}^{z})^{2}\rangle\rangle_{r},
\end{equation}
with the coefficients
\begin{eqnarray}\label{eq14}
\nonumber  a_{1} &=& F(\gamma), \\
\nonumber  a_{2} &=& \mathrm{sinh}(J\nabla)F(x+\gamma)|_{x=0}, \\
a_{3} &=& \mathrm{cosh}(J\nabla)F(x+\gamma)|_{x=0},
\end{eqnarray}
where $\gamma=(q-1)A$ is the effective field produced by the $(q-1)$ spins outside the system and $A$ is an unknown parameter to be determined self-consistently. In the effective-field approximation, the number of independent spin variables describes the considered system. This number is given by the relation $\nu=\langle\langle (S_{i}^{z})^{2S}\rangle\rangle_{r}$. As an example for the spin-1 system, $2S=2$ which means that we have to introduce the additional parameters $\langle\langle (S_{0}^{z})^{2}\rangle\rangle_{r}$ and $\langle\langle (S_{\delta}^{z})^{2}\rangle\rangle_{r}$ resulting from the usage of the Van der Waerden identity for the spin-1 Ising system. With the help of Eq. (\ref{eq5}), quadrupolar moment of the central spin can be obtained as follows
\begin{equation}\label{eq15} \langle\langle
(S_{0}^{z})^{2}\rangle\rangle_{r}=\left\langle\left\langle\prod_{j=1}^{q}\left[1+S_{j}^{z}\mathrm{sinh}(J\nabla)+(S_{j}^{z})^{2}\{\mathrm{cosh}(J\nabla)-1\}\right]\right\rangle\right\rangle_{r}
G(x)|_{x=0},
\end{equation}
where the function $G(x)$ is defined as
\begin{equation}\label{eq16}
G(x)=\int dD_{i}P(D_{i})g(x,D_{i}).
\end{equation}
Definition of the function $g(x,D_{i})$ in Eq. (\ref{eq16}) is given as follows and the expression in Eq. (\ref{eq16a}) can be evaluated by using the eigenvalues and corresponding eigenvectors of the effective Hamiltonian matrix in Eq. (\ref{eq7}).
\begin{eqnarray}\label{eq16a}
g(x,D_{i})&=&\frac{1}{\sum_{n=1}^{3}\exp(\beta\lambda_{n})}\sum_{n=1}^{3}\langle \varphi_{n}|\left(S_{i}^{z}\right)^{2}|\varphi_{n}\rangle\exp(\beta\lambda_{n}),\\
\nonumber
&=&\frac{2\cosh(\beta x)}{2\cosh(\beta x)+e^{-\beta D_{i}}}.
\end{eqnarray}
Hence, we get the quadrupolar moment by expanding the right-hand side of Eq. (\ref{eq15})
\begin{eqnarray}\label{eq17}
\nonumber  \langle\langle (S_{0}^{z})^{2}\rangle\rangle_{r} &=& r_{0}+3n_{1}\langle\langle S_{1}\rangle\rangle_{r}+3(r_{1}-r_{0})\langle\langle S_{1}^{2}\rangle\rangle_{r}+3r_{2}\langle\langle S_{1}S_{2}\rangle\rangle_{r} \\
\nonumber   &&+6(n_{2}-n_{1})\langle\langle S_{1}S_{2}^{2}\rangle\rangle_{r}+3(r_{0}-2r_{1}+r_{3})\langle\langle S_{1}^{2}S_{2}^{2}\rangle\rangle_{r}+n_{3}\langle\langle S_{1}S_{2}S_{3}\rangle\rangle_{r}  \\
\nonumber   &&+ 3(r_{4}-r_{2})\langle\langle S_{1}S_{2}S_{3}^{2}\rangle\rangle_{r}+3(n_{1}-2n_{2}+n_{4})\langle\langle S_{1}S_{2}^{2}S_{3}^{2}\rangle\rangle_{r}  \\
   &&+(-r_{0}+3r_{1}-3r_{3}+r_{5})\langle\langle
   S_{1}^{2}S_{2}^{2}S_{3}^{2}\rangle\rangle_{r},
\end{eqnarray}
with
\begin{eqnarray}\label{eq18}
\nonumber  r_{0}=G(0),\ \ \ \ \ \ \ \ \ \ \ \ \ \ \ \ \ \ \ \ \ \ \ \ \ \ \ \ \ \ \ \ \ \ \ \ \ \ \ \ \  & &  \\
\nonumber  r_{1}= \mathrm{cosh}(J\nabla)G(x)|_{x=0},\ \ \ \ \ \ \ \ \ \ \ \ \ \ \ \ \ \ \ \ \ \ \ & & n_{1}= \mathrm{sinh}(J\nabla)G(x)|_{x=0},  \\
\nonumber  r_{2}= \mathrm{sinh}^{2}(J\nabla)G(x)|_{x=0},\ \ \ \ \ \ \ \ \ \ \ \ \ \ \ \ \ \ \ \ \ \ & & n_{2}= \mathrm{cosh}(J\nabla) \mathrm{sinh}(J\nabla)G(x)|_{x=0},  \\
\nonumber  r_{3}= \mathrm{cosh}^{2}(J\nabla)G(x)|_{x=0}, \ \ \ \ \ \ \ \ \ \ \ \ \ \ \ \ \ \ \ \ \ \ & & n_{3}= \mathrm{sinh}^{3}(J\nabla)G(x)|_{x=0}, \\
\nonumber  r_{4}= \mathrm{cosh}(J\nabla) \mathrm{sinh}^{2}(J\nabla)G(x)|_{x=0}, \ \ \ \ \ \ \ \ \ \ & & n_{4}= \mathrm{cosh}^{2}(J\nabla) \mathrm{sinh}(J\nabla)G(x)|_{x=0}, \\
r_{5}=\mathrm{cosh}^{3}(J\nabla)G(x)|_{x=0}, \ \ \ \
\ \ \ \ \ \ \ \ \ \ \ \ \ \ \ \ \ \ & &
\end{eqnarray}
Corresponding to Eq. (\ref{eq12}),
\begin{equation}\label{eq19} \langle\langle
(S_{\delta}^{z})^{2}\rangle\rangle_{r}=\langle\langle
1+S_{0}^{z}\mathrm{sinh}(J\nabla)+(S_{0}^{z})^{2}\{\mathrm{cosh}(J\nabla)-1\}\rangle\rangle_{r}
G(x+\gamma),
\end{equation}
\begin{equation}\label{eq20}
\langle\langle
S_{1}^{2}\rangle\rangle_{r}=b_{1}\left(1-\langle\langle
(S_{0}^{z})^{2}\rangle\rangle_{r}\right)+b_{2}\langle\langle
S_{0}^{z} \rangle\rangle_{r}+b_{3}\langle\langle
(S_{0}^{z})^{2}\rangle\rangle_{r}.
\end{equation}
where
\begin{eqnarray}\label{eq21}
\nonumber  b_{1} &=& G(\gamma), \\
\nonumber  b_{2} &=& \mathrm{sinh}(J\nabla)G(x+\gamma)|_{x=0}, \\
b_{3} &=& \mathrm{cosh}(J\nabla)\rangle G(x+\gamma)|_{x=0}.
\end{eqnarray}
Eqs. (\ref{eq10}), (\ref{eq13}), (\ref{eq17}) and (\ref{eq20}) are the fundamental spin identities of the system. When the right-hand sides of Eqs. (\ref{eq8}) and (\ref{eq15}) are expanded, the multispin correlation functions appear. The simplest approximation, and one of the most frequently adopted is to decouple these correlations according to
\begin{equation}\label{eq22}
\left\langle\left\langle
S_{i}^{z}(S_{j}^{z})^{2}...S_l^z\right\rangle\right\rangle_{r}\cong\left\langle\left\langle
S_i^z\right\rangle\right\rangle_{r}\left\langle\left\langle
(S_j^z)^{2}\right\rangle\right\rangle_{r}...\left\langle\left\langle
S_l^z\right\rangle\right\rangle_{r},
\end{equation}
for $i\neq j \neq...\neq l$ \cite{tamura_kaneyoshi}. The main difference of the method used in this study from the other approximations in the literature emerges in comparison with any decoupling approximation (DA) when expanding the right-hand sides of Eqs. (\ref{eq8}) and (\ref{eq15}). In other words, one advantage of the approximation method used in this study is that no uncontrolled decoupling procedure is used for the higher-order correlation functions.

For spin-1 Ising system with $q=3$, taking Eqs. (\ref{eq10}), (\ref{eq13}), (\ref{eq17}) and (\ref{eq20}) as a basis, we derive a set of linear equations of the spin identities. At this point, we assume that (i) the correlations depend only on the distance between the spins, (ii) the average values of a central spin and its nearest-neighbor spin (it is labeled as the perimeter spin) are equal to each other with the fact that, in the matrix representations of spin operator $\hat{S}$, the spin-1 system has the properties $(S_{\delta}^{z})^{3}=S_{\delta}^{z}$ and $(S_{\delta}^{z})^{4}=(S_{\delta}^{z})^{2}$. Thus, the number of the set of linear equations obtained for the spin-1 Ising system with $q=3$ reduces to twenty one, and the complete set is given in Appendix \ref{appendixb}.

If Eq. (\ref{eqB1}) is written in the form of a $21\times21$ matrix and solved in terms of the variables $x_{i}[(i=1,2,...,21)(e.g., x_{1}=\langle\langle S_{0}^{z}\rangle\rangle_{r}, x_{2}=\langle\langle S_{1}S_{0}\rangle\rangle_{r},...)]$ of the linear equations, all of the spin correlation functions, as well as magnetizations and quadrupolar moments can be easily determined as functions of the temperature and Hamiltonian parameters. Since the thermal and configurational average of the central spin is equal to that of its nearest-neighbor spins within the present method then the unknown parameter $A$ can be numerically determined by the relation
\begin{equation}\label{eq23}
\langle\langle S_{0}^{z}\rangle\rangle_{r}=\langle\langle
S_{1}\rangle\rangle_{r} \qquad {\rm{ or }}\qquad x_{1}=x_{4}.
\end{equation}
By solving Eq. (\ref{eq23}) numerically at a given fixed set of Hamiltonian parameters we obtain the parameter $A$. Then we use the numerical values of $A$ to obtain the spin correlation functions which can be found from Eq. (\ref{eqB1}). Note that $A=0$ is always the root of Eq. (\ref{eq23}) corresponding to the disordered state of the system. The nonzero root of $A$ in Eq. (\ref{eq23}) corresponds to the long-range ordered state of the system. Once the spin identities have been evaluated then we can give the numerical results for the thermal and magnetic properties of the system. Since the effective field $\gamma$ is very small in the vicinity of $k_{B}T_{c}/J$, we can obtain the critical temperature for the fixed set of Hamiltonian parameters by solving Eq. (\ref{eq23}) in the limit of $\gamma\rightarrow0$ then we can construct the whole phase diagrams of the system. Depending on the values of Hamiltonian and crystal field distribution parameters, there may be two solutions [i.e., two critical temperature values which satisfy Eq. (\ref{eq23})] corresponding to the first (or second) and second-order phase-transition points, respectively. We determine the type of the transition by looking at the temperature dependence of magnetization for selected values of system parameters.

\section{Results and Discussion}\label{results}
In this section, we will discuss the effect of the crystal field distributions defined in Eqs. (\ref{eq2}) and (\ref{eq3}) on the global phase diagrams of the system where the second and first order transitions are shown by solid and dashed curves, respectively with tricritical points (shown by hollow circles) for honeycomb $(q=3)$ and square $(q=4)$ lattices. Also, in order to clarify the type of the transitions in the system, we will give the temperature dependence of the order parameter.

\subsection{Phase diagrams of the system with dilute crystal field}
\begin{figure}
\includegraphics[width=8.9cm]{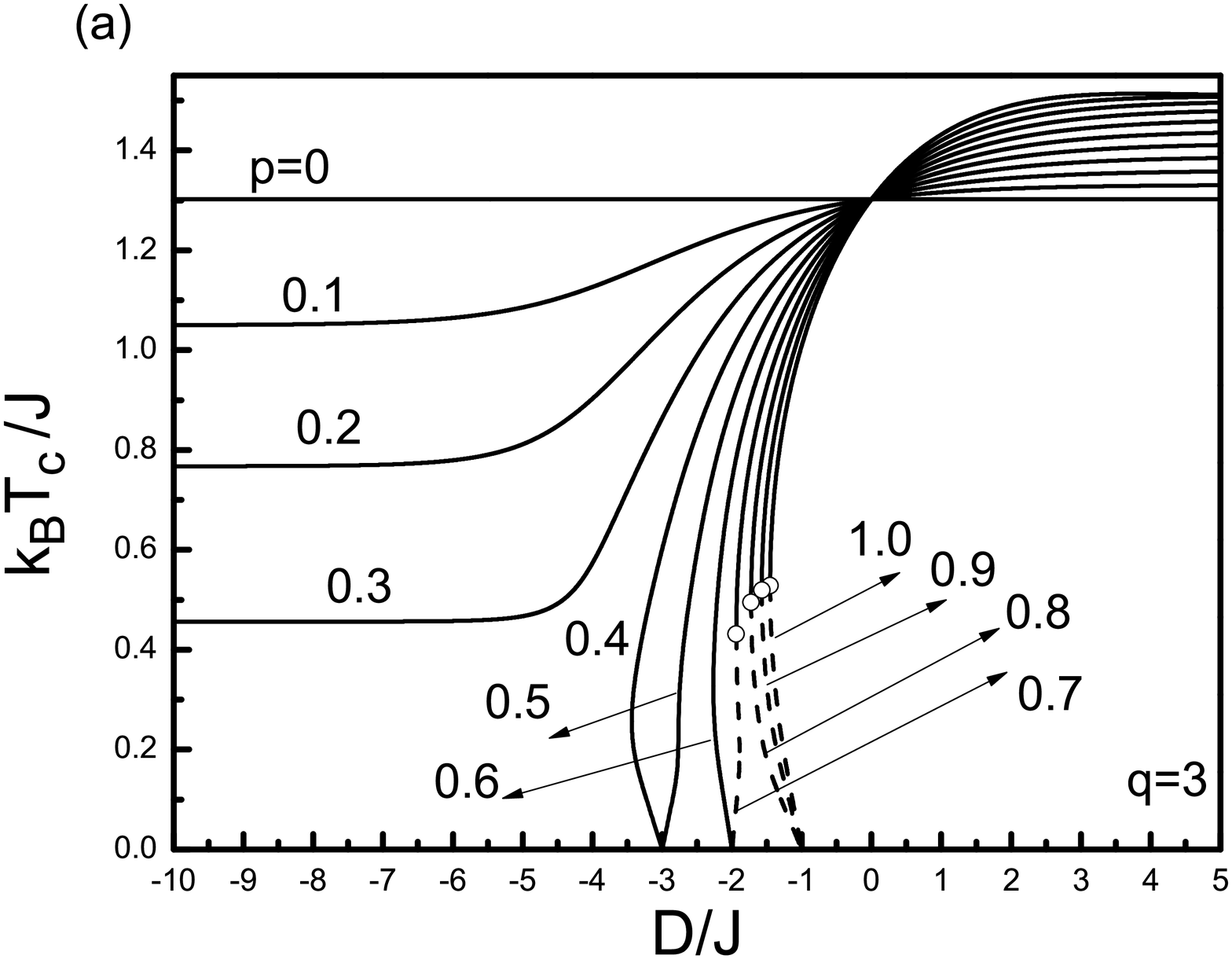}
\includegraphics[width=8.9cm]{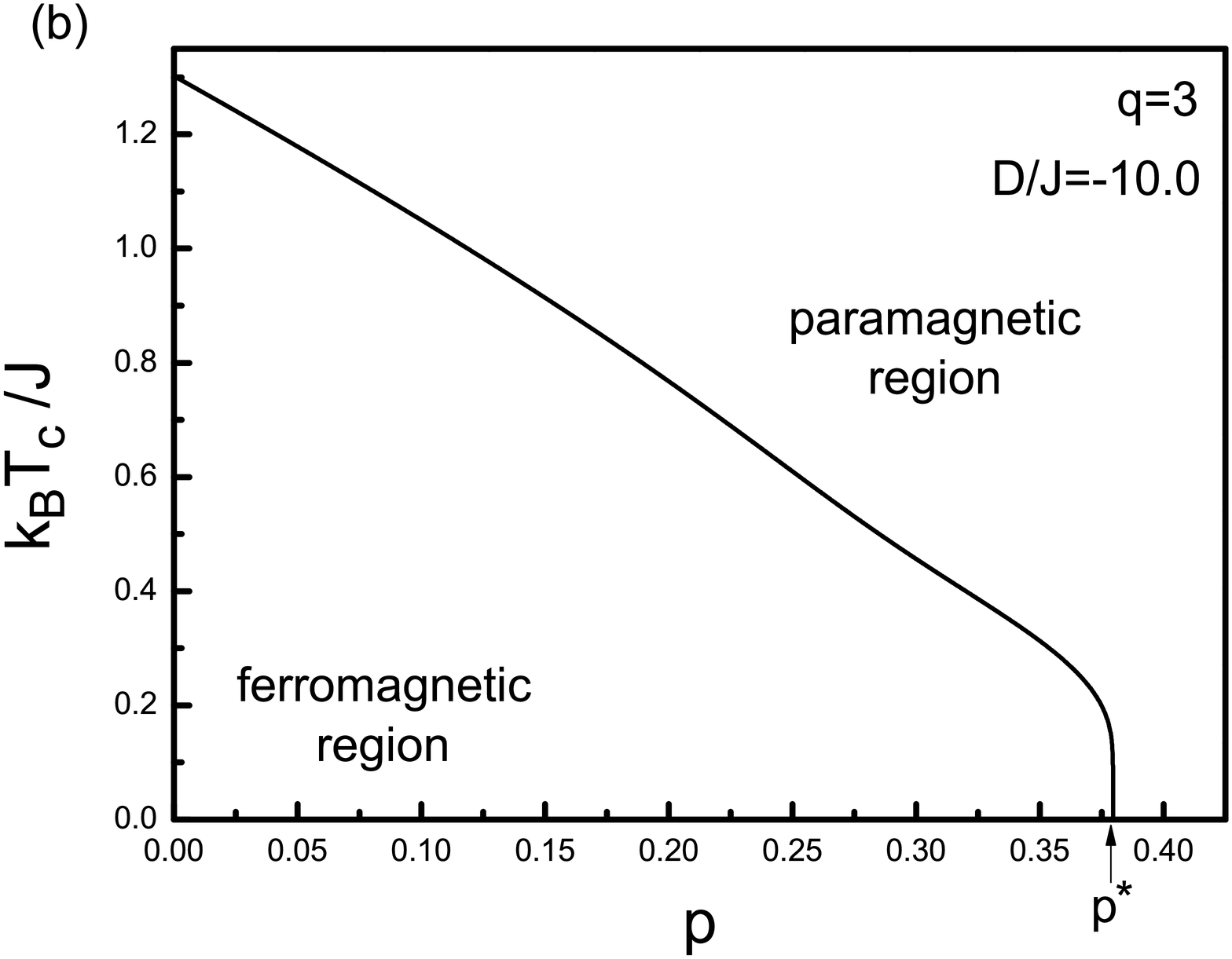}\\
\includegraphics[width=8.9cm]{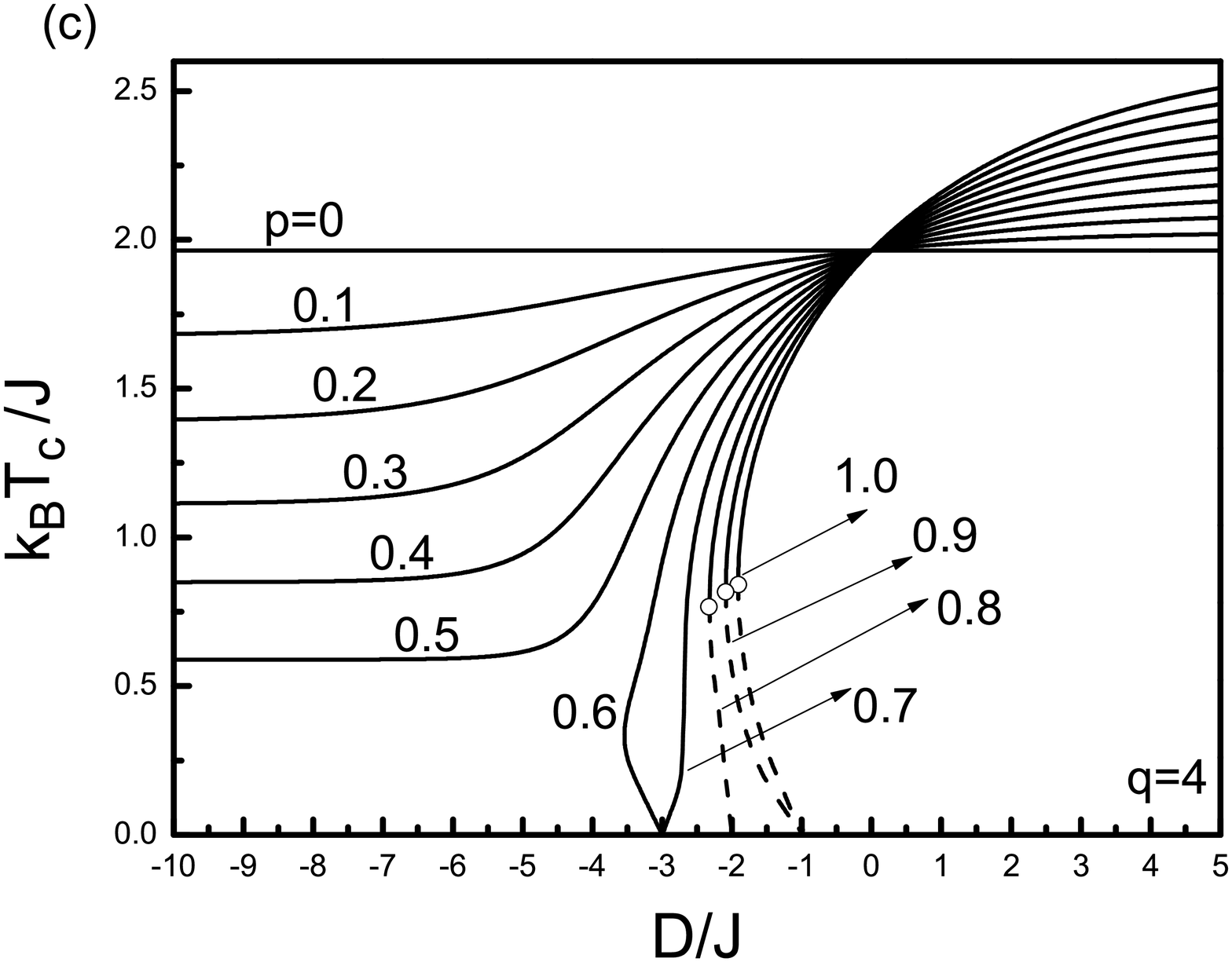}
\includegraphics[width=8.9cm]{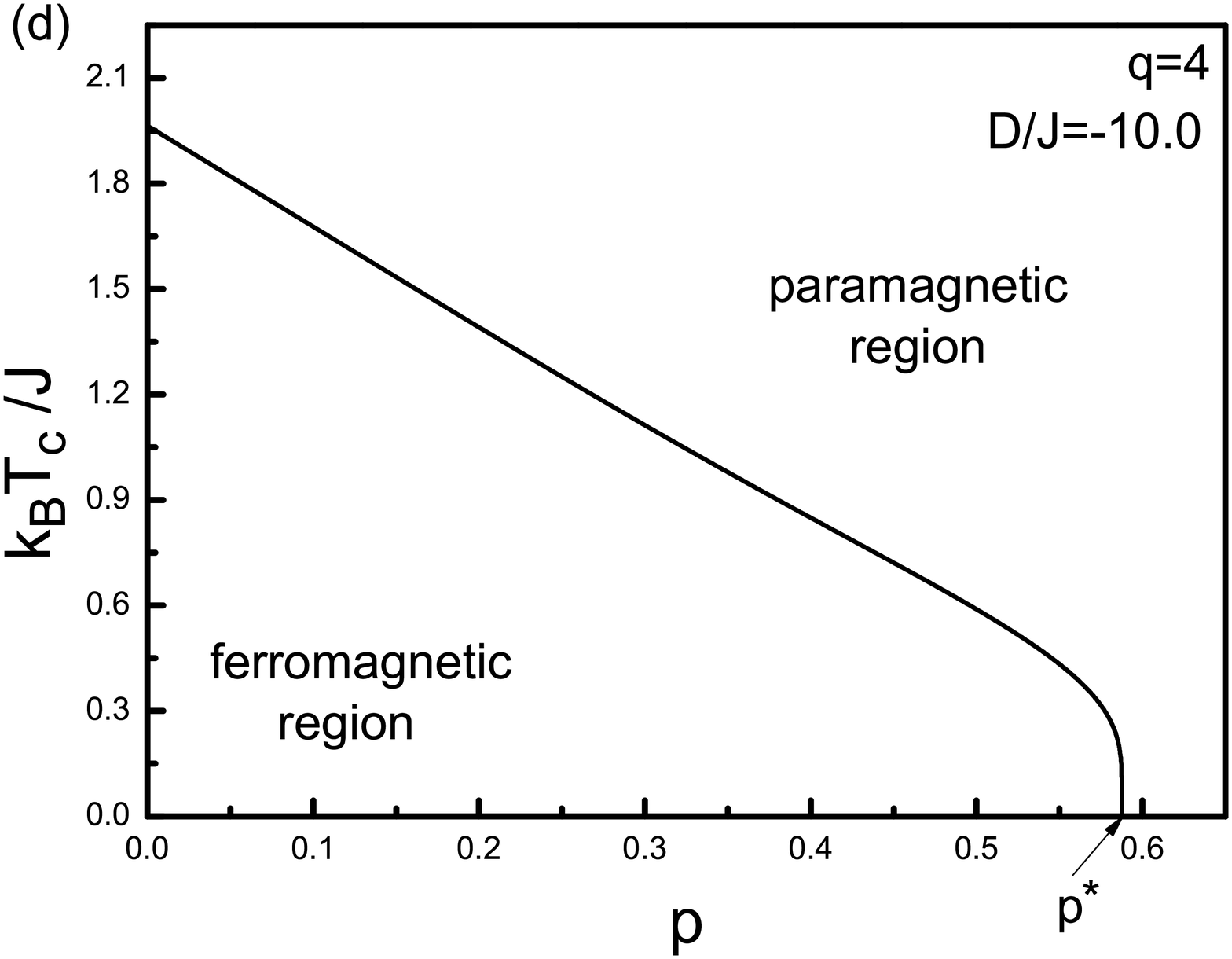}\\
\caption{(a) Phase diagrams of the system for $q=3$ in a $(k_{B}T_{c}/J-D/J)$ plane corresponding to dilute crystal field distribution defined in Eq. (\ref{eq2}). The solid and dashed lines correspond to second- and first-order phase transitions, respectively. The open circles denote the tricritical points, and the numbers on each curve represent the value of concentration $p$. (b) Phase diagrams of the system for $q=3$ in a $(k_{B}T_{c}/J-p)$ plane with a selected value of the crystal field $D/J=-10.0$. (c) Phase diagrams of the system for $q=4$ in a $(k_{B}T_{c}/J-D/J)$ plane corresponding to dilute crystal field distribution defined in Eq. (\ref{eq2}). The solid and dashed lines correspond to second- and first-order phase transition, respectively. The open circles refer to the tricritical points, and the numbers on each curve represent the value of concentration $p$. (d) Phase diagrams of the system for $q=4$ in a $(k_{B}T_{c}/J-p)$ plane with a selected value of the crystal field $D/J=-10.0$.}\label{fig1}
\end{figure}

In this section, we illustrate the phase diagrams and magnetization curves of the system with a dilute crystal field distribution defined in Eq.(\ref{eq2}) where crystal field $D$ is turned on, or turned off with probabilities $p$ and $(1-p)$ on the lattice sites, respectively. In Figs. (\ref{fig1}a) and (\ref{fig1}c), we plot the phase diagrams of the system in $(k_{B}T_{c}/J-D/J)$ plane for honeycomb and square lattices with coordination numbers $q=3$ and $q=4$, respectively. As seen in Figs. (\ref{fig1}a) and (\ref{fig1}c), phase diagrams of the system can be divided into three parts with different concentration values $p$. For the curves in the first group with $p<p^{*}$, the system always exhibits a second order phase transition with a finite critical temperature $k_{B}T_{c}/J$ which extent to $D/J\rightarrow-\infty$. If the concentration $p$ reaches its critical value $p^{*}$ then the critical temperature depresses to zero. Physical reason underlying this behavior can be explained as follows; when we select sufficiently large negative crystal field values $(\mathrm{i.e. } D/J\rightarrow-\infty)$, all of the spins in the system tend to align in $S=0$ state. As $p$ increases starting from zero, the ratio of spins which aligned in $S=0$ state increases, and therefore, magnetization weakens, and accordingly, critical temperature of the system decreases. According to our numerical results, the critical concentration value is obtained as $p^{*}=0.3795$ for $q=3$ and $p^{*}=0.5875$ for $q=4$. In the second group of the phase diagrams in Figs. (\ref{fig1}a) and (\ref{fig1}c), the system exhibits a reentrant behavior of second order, whereas the curves in the third group, exhibit a reentrant behavior of first order with a tricritical point at which a first order transition line turns into a second order transition line. Besides, the curves which exhibit a reentrant behavior of first (or second) order, depress to zero at three successive values of crystal field $D/J=-3.0, -2.0, -1.0$. Moreover, in $D/J\rightarrow\infty$ limit, the system behaves like spin-$1/2$ for $p=1.0$. In the case of $p\neq0$, the ratio of spins that behave like $S=\pm1$ increases as $p$ increases. Therefore, for $0\leq p\leq 1.0$, all transition lines have finite critical temperatures which increase with increasing $p$ values for $D/J\rightarrow \infty$. At this point, we also note that if we select $D/J=0$ in Eq. (\ref{eq2}), all lattice sites expose to a crystal field $D_{i}/J=0$ independent from $p$. Hence, all transition lines intersect each other on the point $(D/J,k_{B}T_{c}/J)=(0,1.3022)$ for $q=3$, and $(D/J,k_{B}T_{c}/J)=(0,1.9643)$ for $q=4$. Meanwhile, previous studies based on EFT are not capable of obtaining first order transition lines of the system. From this point of view, we see that our method improves the results of the other EFT works and we take the conventional EFT method one step forward by investigating the global phase diagrams, especially the first-order transition lines that include reentrant phase transition regions.
\begin{table}[!h]
\begin{center}
\begin{threeparttable}
\caption{Critical concentration $p^{*}$ obtained by several methods and the present work for honeycomb $(q=3)$ and square $(q=4)$ lattices.}
\renewcommand{\arraystretch}{1.3}
\begin{tabular}{llllllllllll}
\thickhline
Lattice & EFT-I \cite{kaneyoshi3,kaneyoshi4} & \ \ \ EFT-II \cite{yan} & \ \ \ MFT\cite{benyoussef2,carneiro2} & \ \ PA \cite{lara}  & \ \ Present Work\\
\hline $q=3$  & \ 0.484 & \ \ \ \ 0.492 & \ \ \ \ 1.0 & \ \ 0.5  & \ \ \ \ \ \ \ 0.3795  \\
$q=4$   & \ 0.604 & \ \ \ \ 0.610 & \ \ \ \ 1.0 & \ \ 0.667  & \ \ \ \ \ \ \ 0.5875 \\
\thickhline \\
\end{tabular}\label{table1}
\end{threeparttable}
\end{center}
\end{table}

On the other hand, Figs. (\ref{fig1}b) and (\ref{fig1}d) shows the phase boundary in $(k_{B}T_{c}/J-p)$ plane which separates the ferromagnetic and paramagnetic phases with $D/J\rightarrow-\infty$. According to this figure, critical temperature $k_{B}T_{c}/J$ of system decreases gradually, and ferromagnetic region gets narrower as $p$ increases, and $k_{B}T_{c}/J$ value depresses to zero at $p=p^{*}$. Such a behavior is an expected fact in dilution problems. Numerical value of critical concentration $p^{*}$ for honeycomb $(q=3)$ and square $(q=4)$ lattices is given in Table \ref{table1}, and compared with the other works in the literature. As seen in Table \ref{table1}, numerical values of $p_{c}$ for $q=3$ and $q=4$ are new results in literature. Furthermore, MFT \cite{benyoussef2,carneiro2} predicts that the system always has a finite critical temperature and exists in a ferromagnetic state at lower temperatures in $D/J\rightarrow-\infty$ limit, except that $p=1.0$. This artificial result can be regarded as a failure of the MFT.

\begin{figure}
\includegraphics[width=14cm]{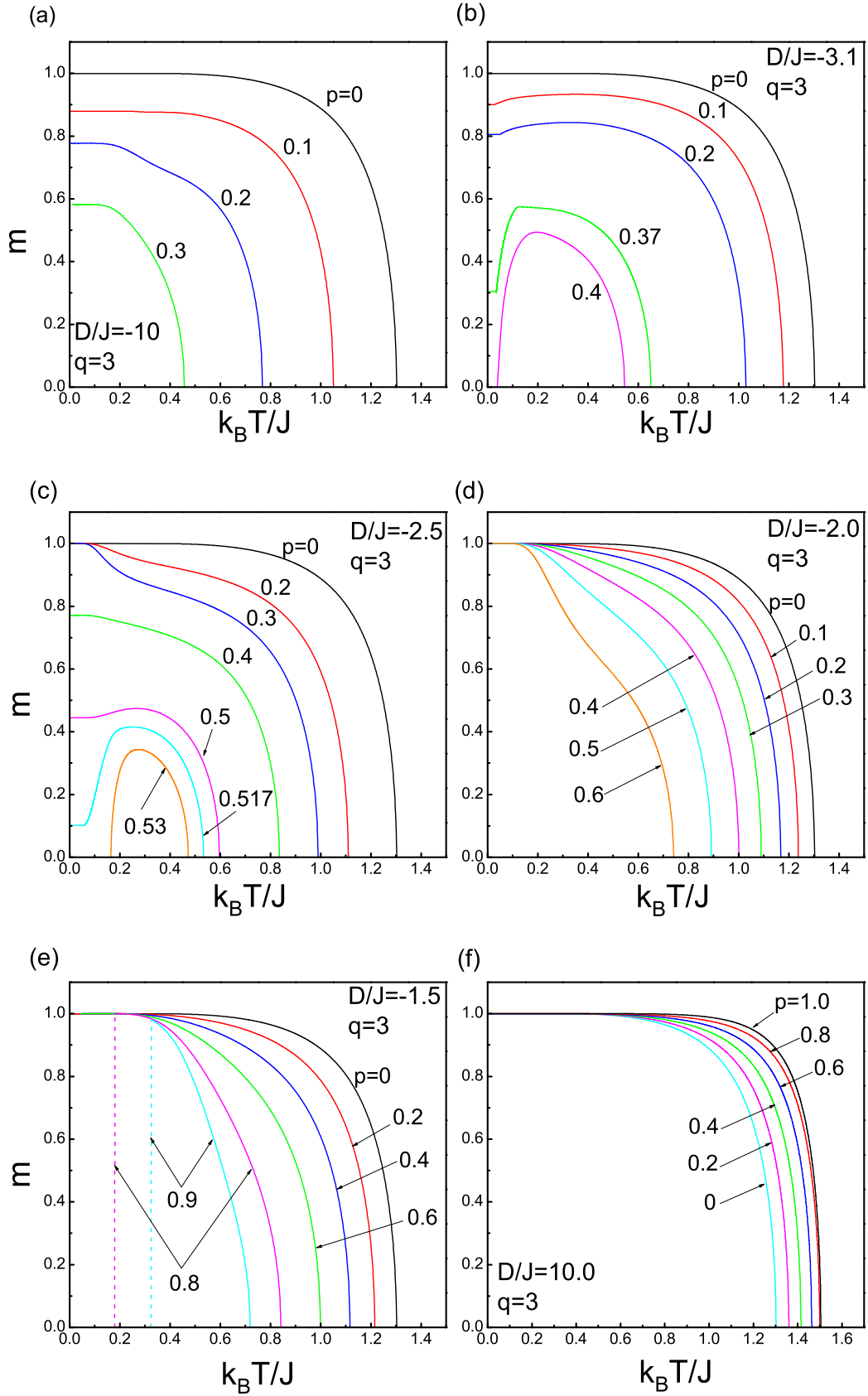}\\
\caption{(Color online) Temperature dependence of magnetization corresponding to Fig. (\ref{fig1}a) with some selected values of crystal field. (a) $D/J=-10.0$, (b) $D/J=-3.1$, (c) $D/J=-2.5$, (d) $D/J=-2.0$, (e) $D/J=-1.5$, and (f) $D/J=10.0$. The numbers on each curve denote the value of concentration $p$. The solid and dashed lines correspond to second- and first-order phase transitions, respectively.}\label{fig2}
\end{figure}

In Fig.(\ref{fig2}), we plot the temperature dependencies of magnetization curves corresponding to the phase diagrams depicted in Fig. (\ref{fig1}) for $q=3$. As seen in Fig. (\ref{fig2}), as $p$ increases then critical temperature $k_{B}T_{c}/J$ values decrease for $D/J<0$, except the reentrant phase transition temperatures which occur at low temperatures. On the other hand, effect of increasing $p$ values on the shape of magnetization curves depends on value of $D/J$. Namely, in Figs. (\ref{fig2}a) and (\ref{fig2}b) we see that ground state saturation values of magnetization curves decreases as $p$ increases for $D/J=-10.0$ and $-3.1$. Moreover, for $D/J=-3.1$, magnetization curves of the system exhibit a broad maximum at low temperatures for $p=0.37$, and a reentrant behavior of second order for $p=0.4$. If we select $D/J=-2.5$ as in Fig. (\ref{fig2}c), saturation values of magnetization curves remain unchanged for $p=0, 0.2, 0.3$ and tend to decrease for $p>0.3$. If $p$ increases further, a reentrant behavior of second order appears for $p=0.53$, and we see a broad maximum at low temperatures for $p=0.517$ and $0.5$ which tends to depress as $p$ decreases. This broad maximum behavior of magnetization curves originates from the increase in the number of spins directed parallel to the z-direction with increasing temperature, due to the thermal agitation. For $D/J=-2.0$ in Fig. (\ref{fig2}d), magnetization curves saturates at $m=1$ at the ground state and reentrant behavior disappears. If we increase $D/J$ further, for example for $D/J=-1.5$ (Fig. (\ref{fig2}e)), another type of reentrant behavior occurs in the system in which a first order transition is followed by a second order transition. Finally, for sufficiently large positive values of crystal field, magnetization curves always saturate at $m=1$ and the system always undergoes a second order phase transition from a ferromagnetic phase to a paramagnetic phase with increasing temperature, which can be seen in Fig. (\ref{fig2}f) with $D/J=10.0$. As a common property of the curves in Fig. (\ref{fig2}), we see that effect of $p$ on the saturation values, as well as temperature dependence of magnetization curves strictly depend on the strength of $D/J$. Hence, according to us, the presence of dilute crystal fields on the system should produce a competition effect on the phase diagrams of the system. We also note that, although it has not been shown in the present work, magnetization curves for $q=4$ corresponding to the phase diagrams depicted in Fig. (\ref{fig1}c) exhibit qualitatively similar behavior with those of Fig. (\ref{fig2}) with $q=3$.

\begin{figure}
\includegraphics[width=8.9cm]{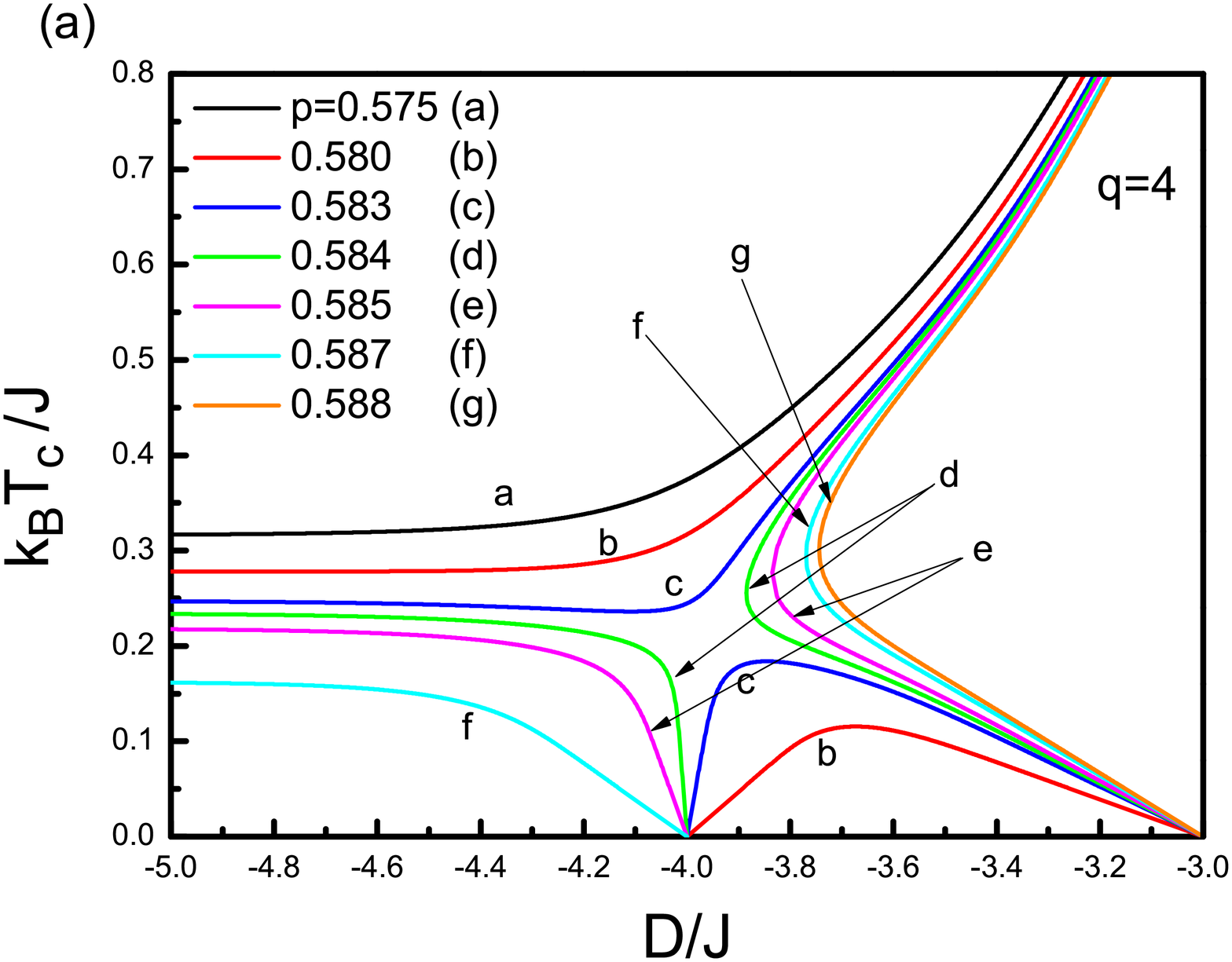}
\includegraphics[width=8.9cm]{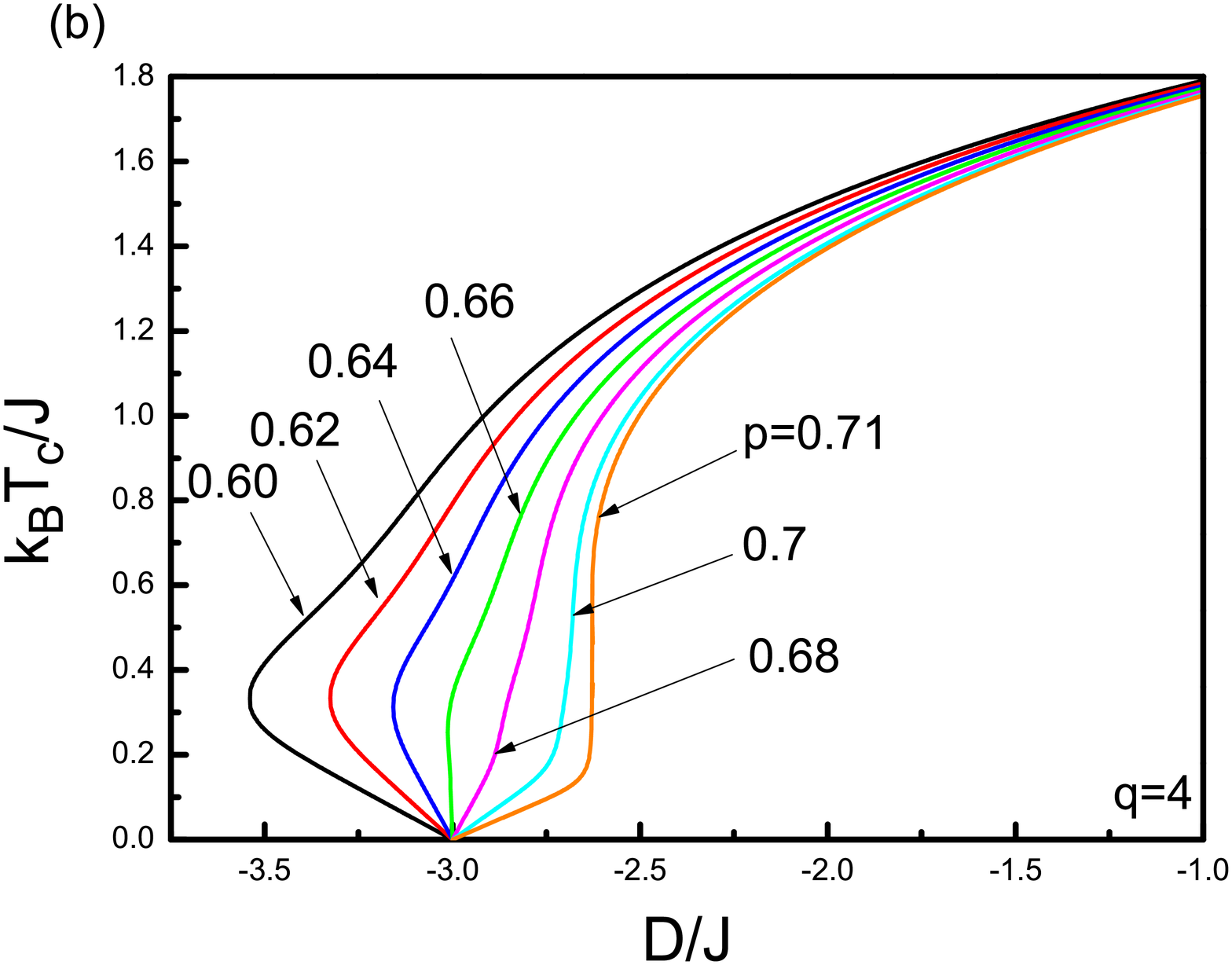}\\
\includegraphics[width=8.9cm]{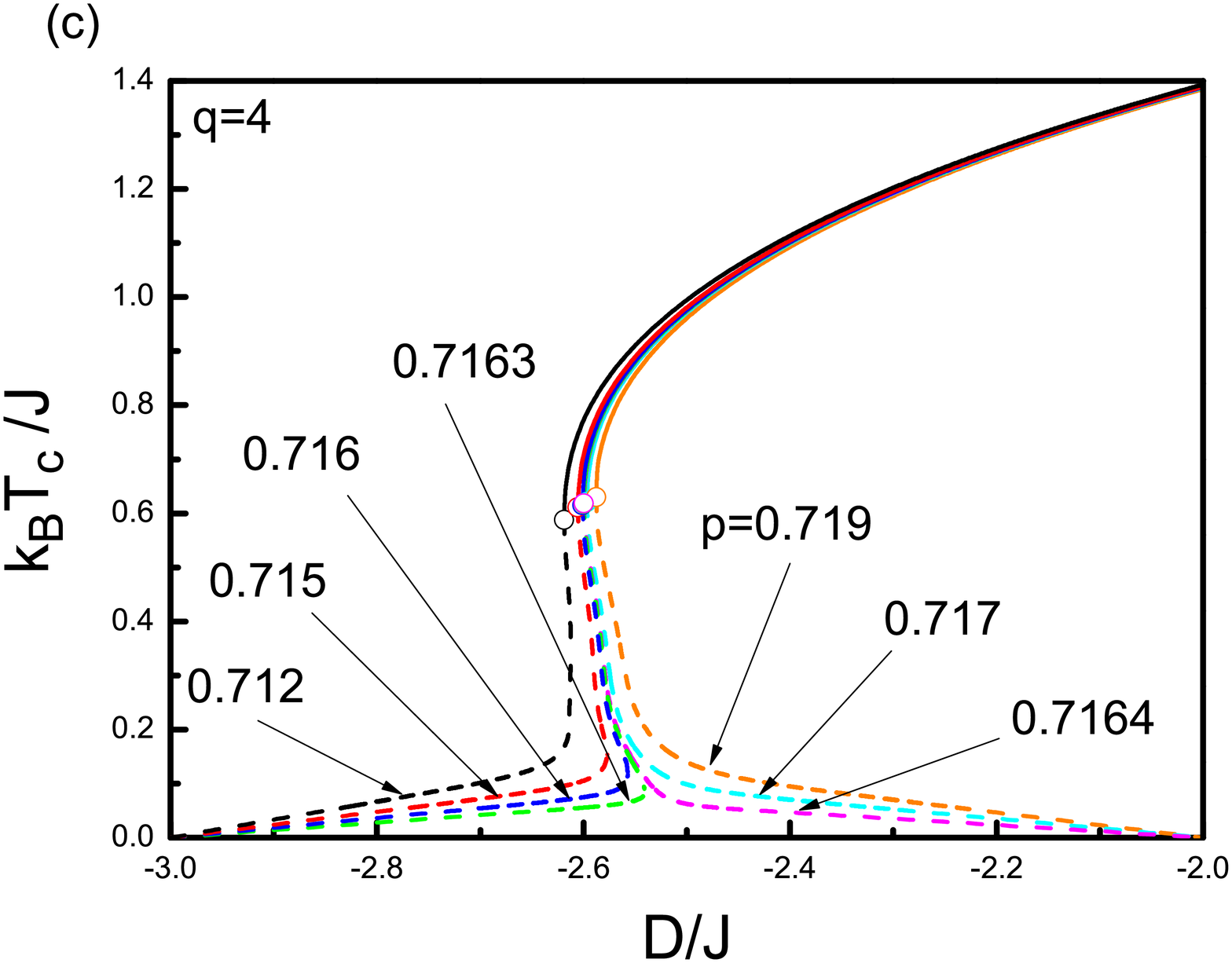}
\includegraphics[width=8.9cm]{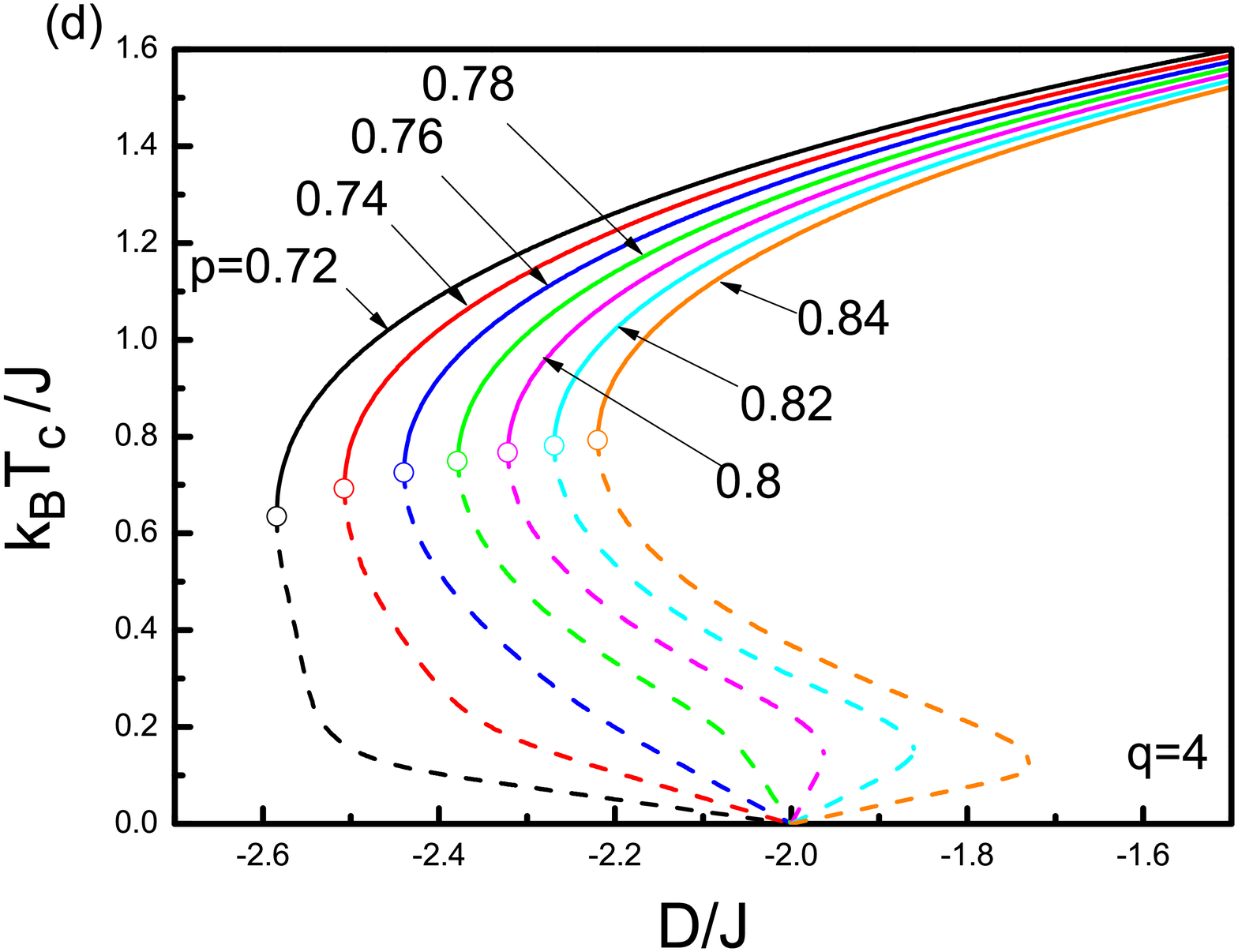}\\
\caption{(Color online) Evolution of the phase diagrams corresponding to Fig. (\ref{fig1}c). The numbers on each curve denote the value of concentration $p$. The solid and dashed lines correspond to second- and first-order phase transitions, respectively. The open circles indicate the tricritical points. }\label{fig3}
\end{figure}

As seen in Fig. (\ref{fig1}), for a dilute crystal field distribution defined in Eq. (\ref{eq2}), the global phase diagrams which are plotted in $(k_{B}T_{c}/J-D/J)$ plane, as well as the phase boundaries in $(k_{B}T_{c}/J-p)$ plane for $q=3$ exhibit qualitatively similar characteristics when compared with those for $q=4$. Hence, in order to examine the phase diagrams which are plotted in $(k_{B}T_{c}/J-D/J)$ plane in Figs. (\ref{fig1}a) and (\ref{fig1}c) in detail, we plot the evolution of the global phase diagrams in Fig. (\ref{fig3}) only for $q=4$. From this point of view, Fig. (\ref{fig3}a) shows how the phase diagrams in Fig. (\ref{fig1}c) evolve when the concentration $p$ changes from 0.5 to 0.6. As seen in Fig. (\ref{fig3}a), we observe a second order phase transition line with a finite critical temperature $k_{B}T_{c}/J$ which extent to $D/J\rightarrow-\infty$ for $p=0.575$. If $p$ increases, namely for $p=0.580$ and $0.583$, we see that a low temperature transition line arises between $-4.0<D/J<-3.0$, as well as a high temperature phase boundary which extents to $D/J\rightarrow-\infty$. If $p$ increases further, such as for $p=0.584$, $0.585$ and $0.587$, high temperature phase boundary is gradually connected to the transition line which arises between $-4.0<D/J<-3.0$, and the phase diagrams exhibit a bulge on the right hand side of $(k_{B}T_{c}/J-D/J)$ plane, whereas another transition line emerges within the range of $-\infty<D/J<-4.0$, which disappears for $p>0.587$. Similarly, evolution of the phase diagrams in Fig. (\ref{fig1}c) when the concentration $p$ changes from 0.6 to 0.7 can be seen in Fig. (\ref{fig3}b). As seen in this figure, the curves for $p=0.60$, $0.62$, $0.64$, $0.66$ exhibit a reentrant behavior of second order, while for $p=0.68$ reentrance disappears and for $p=0.70$ and $0.71$ double reentrance with three successive second order phase transitions occurs in a very narrow region of $D/J$. On the other hand, increasing values of $p$ generates first order phase transitions with tricritical points, as well as reentrant behavior of first order. This phenomena is illustrated in Fig. (\ref{fig3}c). From Fig. (\ref{fig3}c), we see that, the second order transition temperatures decrease as absolute value of $D/J$  increases, and turn into first order transition lines at tricritical points. Evidently, the phase diagrams change abruptly for $p\geq 0.7164$. Hence, the behavior of the $p=0.7163$ curve is completely different from that of $p=0.7164$. Namely, first order transition temperatures of the system for $p\leq 0.7163$ and $p\geq0.7164$ depress to zero at $D/J=-3.0$ and $D/J=-2.0$, respectively. In order to investigate the phase transition features of the system further, we should continue increasing the value of $p$. In Fig. (\ref{fig3}d), we see that the curves for $p=0.72$, $0.74$, $0.76$ and $0.78$ exhibit a reentrant behavior of first order, whereas the curves with $p=0.80$, $0.82$, and $0.84$ exhibit double reentrance with two first order and a second order transition temperature.

\subsection{Phase diagrams of the system with random crystal field}
\begin{figure}
\includegraphics[width=14cm]{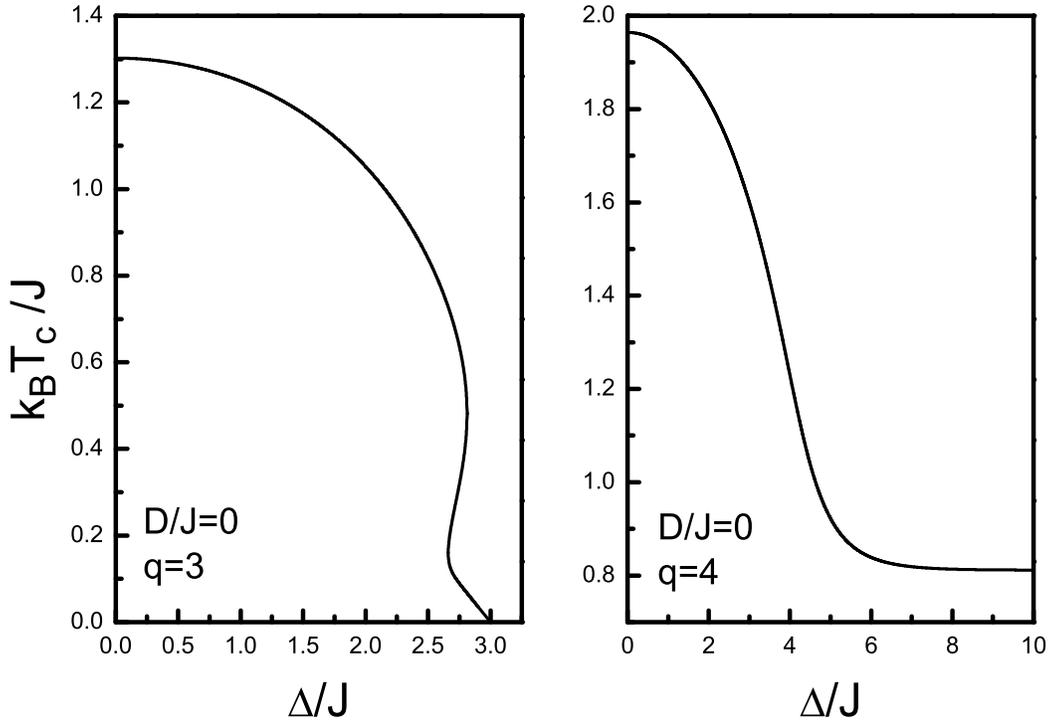}\\
\caption{Phase diagrams of the system in a $(k_{B}T_{c}/J-\triangle/J)$ plane for a bimodal crystal field distribution corresponding to Eq. (\ref{eq3}) with $D/J=0.0$. Left and right-hand side panels are plotted for $q=3$ and $q=4$, respectively.}\label{fig4}
\end{figure}

Next, in order to investigate the effect of the random crystal fields defined in Eq.(\ref{eq3}) on the thermal and magnetic properties of the system, we represent the phase diagrams and corresponding magnetization curves for honeycomb $(q=3)$ and square lattices $(q=4)$ throughout Figs. (\ref{fig4}) and (\ref{fig8}).

We note that random crystal field distribution given in Eq. (\ref{eq3}) with $D/J=0$ corresponds to a bimodal distribution function, while for $\triangle/J=0$, we obtain pure BC model with homogenous crystal field $D/J$. In Fig. (\ref{fig4}), phase diagrams of the system corresponding to the bimodal distribution function are shown in $(k_{B}T_{c}/J-\triangle/J)$ plane. For a bimodal distribution, the phase diagrams have symmetric shape with respect to $\triangle/J$ which comes from the fact that $p=1/2$, and as seen in Fig. (\ref{fig4}), transition temperatures are second order, and it is clear that the system exhibit different characteristic features depending on the coordination number $q$. Namely, for $q=3$, transition temperature decreases with increasing $\triangle/J$ and exhibits double reentrance with three second order phase transition temperatures, then falls to zero at $\triangle/J=3.0$ (left panel in Fig. (\ref{fig4})). On the other hand, as seen on the right panel in Fig. (\ref{fig4}), as $\triangle/J$ increases then the transition temperature of the system for $q=4$ decreases and remains at a finite value for $\triangle/J\rightarrow\infty$ which means that ferromagnetic exchange interactions for $q=3$ are insufficient for the system to keep its ferromagnetic order for $\triangle/J>3.0$, while for $q=4$ these interactions are dominant in the system, and the presence of a disorder in the crystal fields cannot destruct the ferromagnetic order.

\begin{figure}
\includegraphics[width=8.9cm]{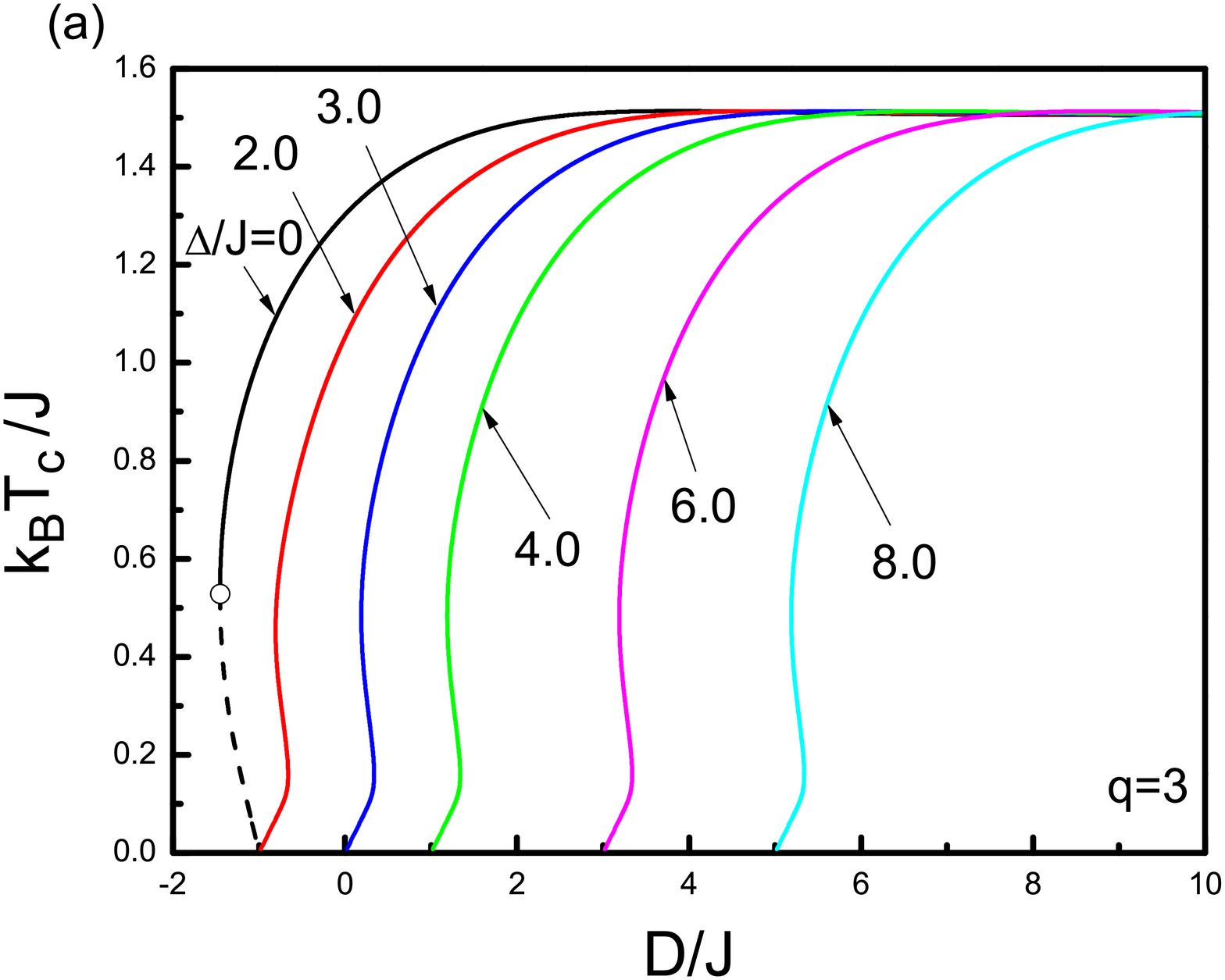}
\includegraphics[width=8.9cm]{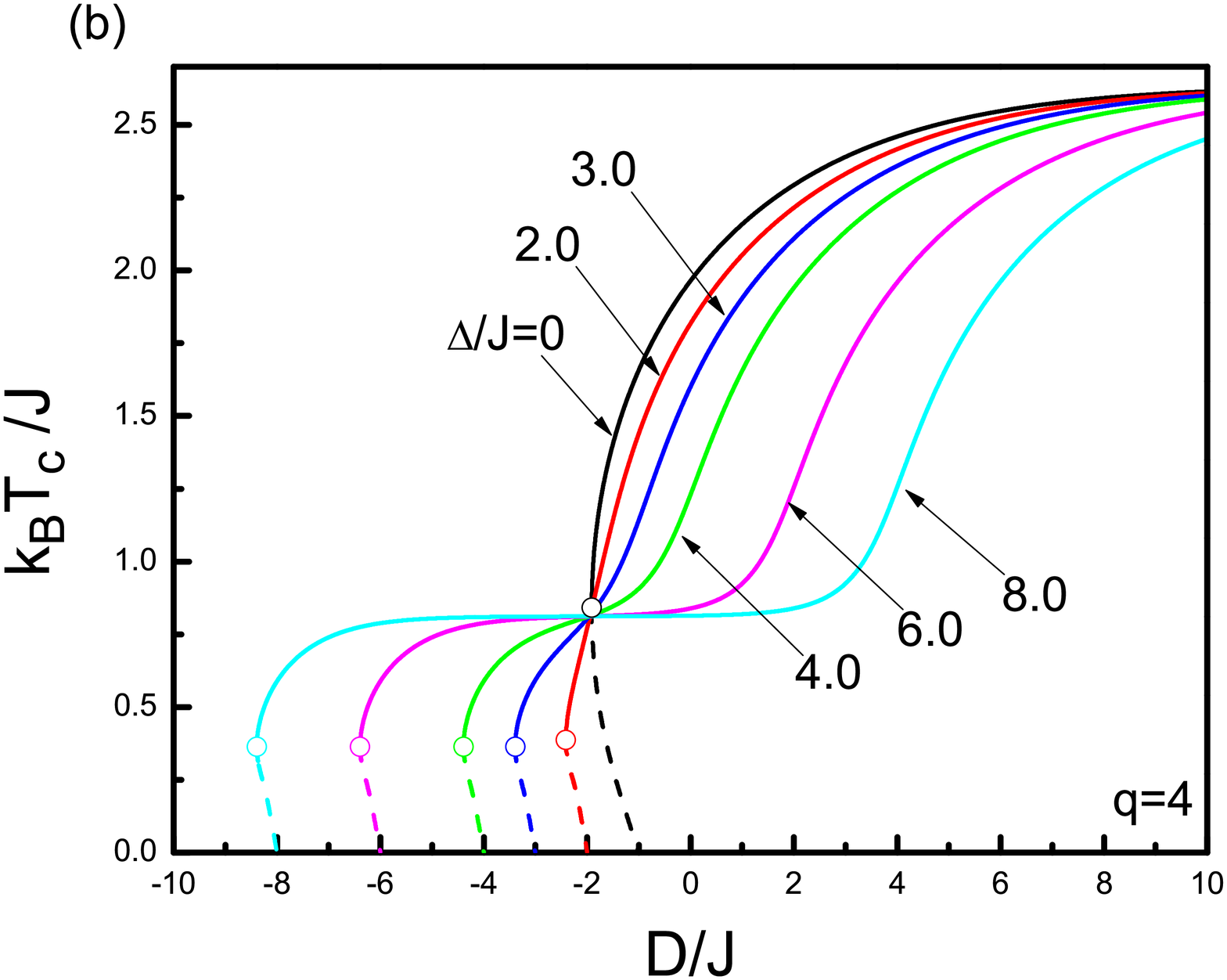}\\
\caption{(Color online) Phase diagrams of the system in a $(k_{B}T_{c}/J-D/J)$ plane corresponding to random crystal field distribution defined in Eq. (\ref{eq3}) for (a) $q=3$, (b) $q=4$. The numbers on each curve denote the value of $\triangle/J$. The open circles represent the tricritical points, and the solid and dashed lines correspond to second- and first-order phase transitions, respectively.  }\label{fig5}
\end{figure}

\begin{figure}
\includegraphics[width=8.9cm]{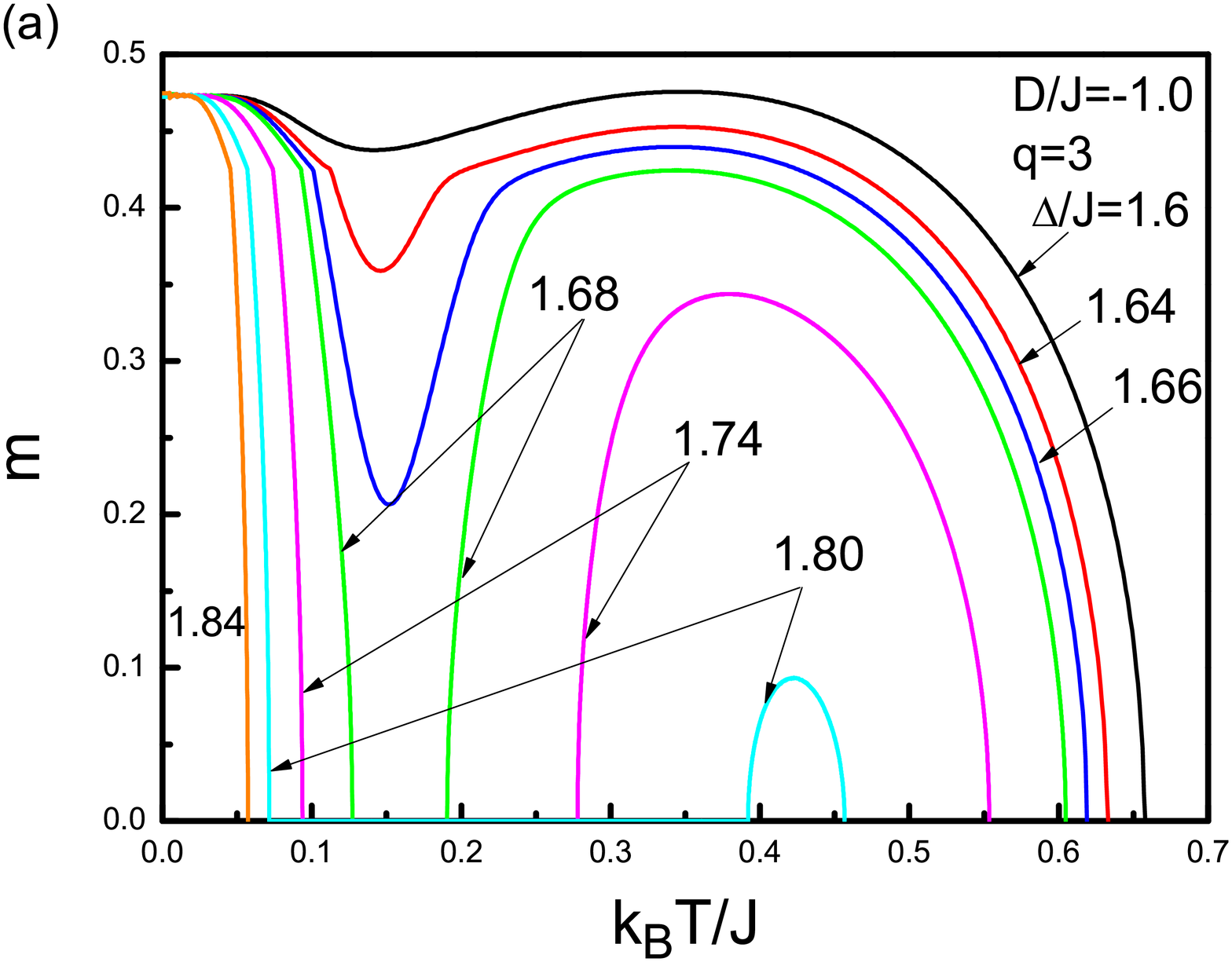}
\includegraphics[width=8.9cm]{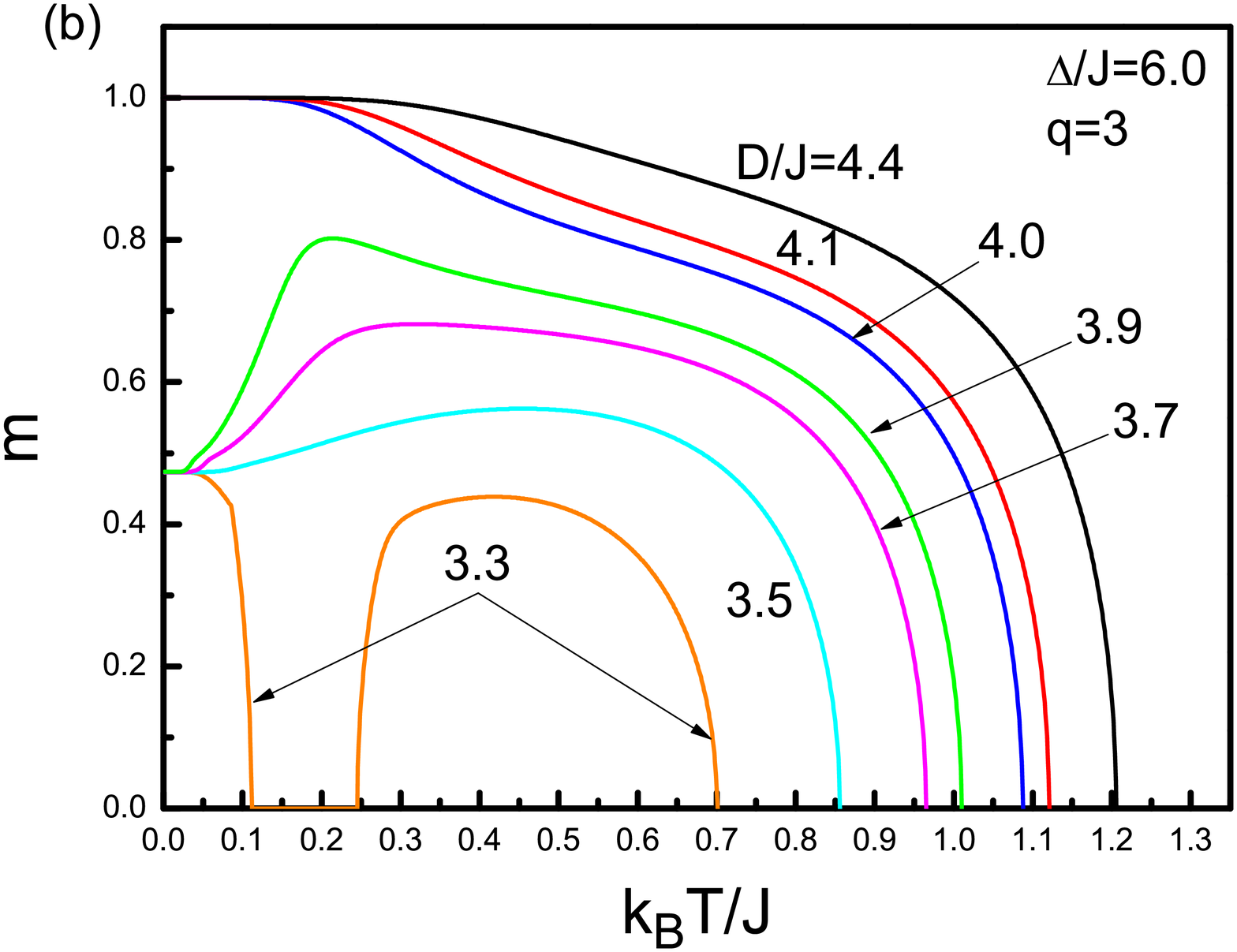}\\
\caption{(Color online) (a) Temperature dependence of magnetization curves corresponding to Fig. (\ref{fig5}a) for $q=3$ (a) with $D/J=-1.0$ and for some selected values of $\triangle/J$, and (b) with $\triangle/J=6.0$ and for some selected values of $D/J$.}\label{fig6}
\end{figure}

\begin{figure}
\includegraphics[width=8.9cm]{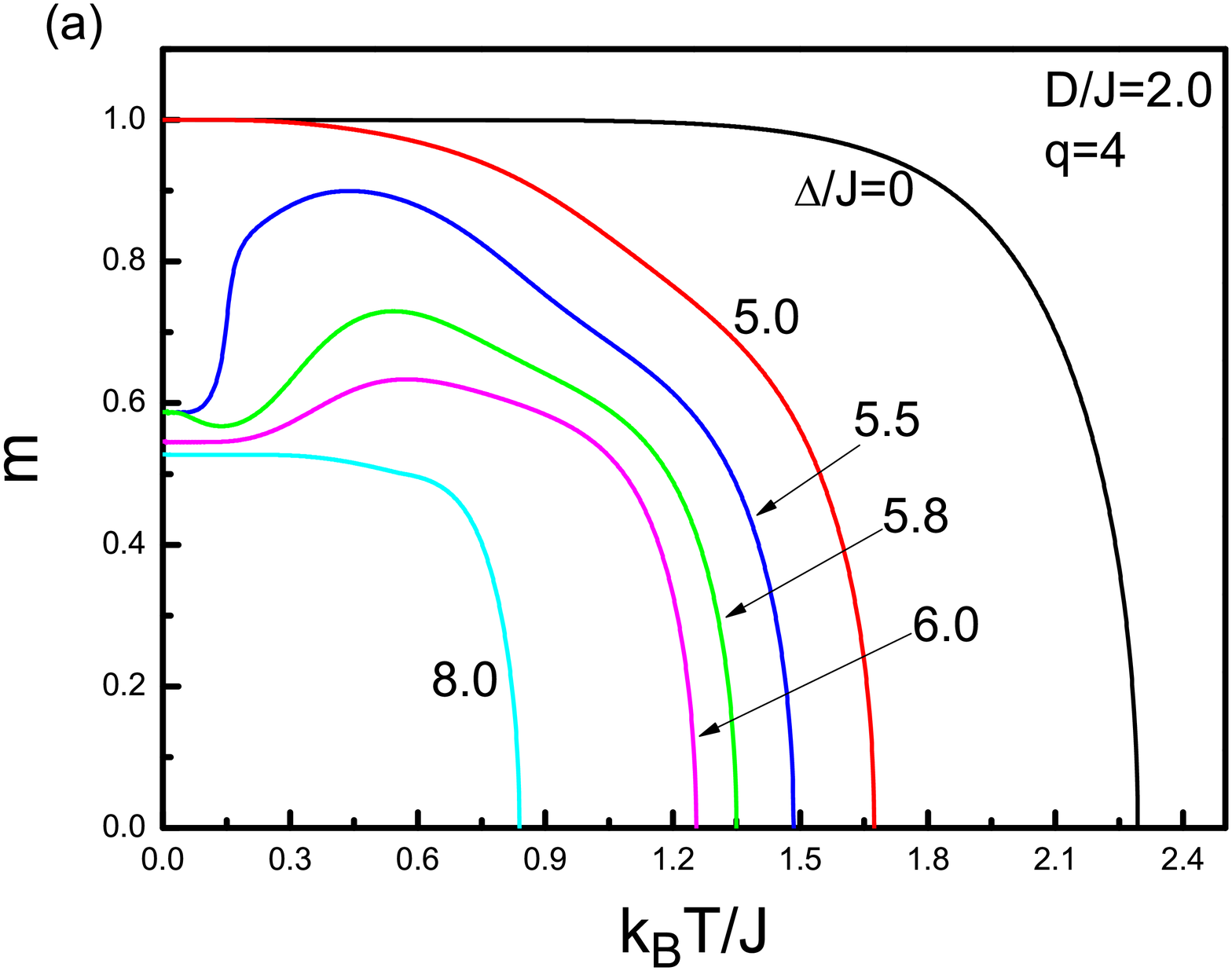}
\includegraphics[width=8.9cm]{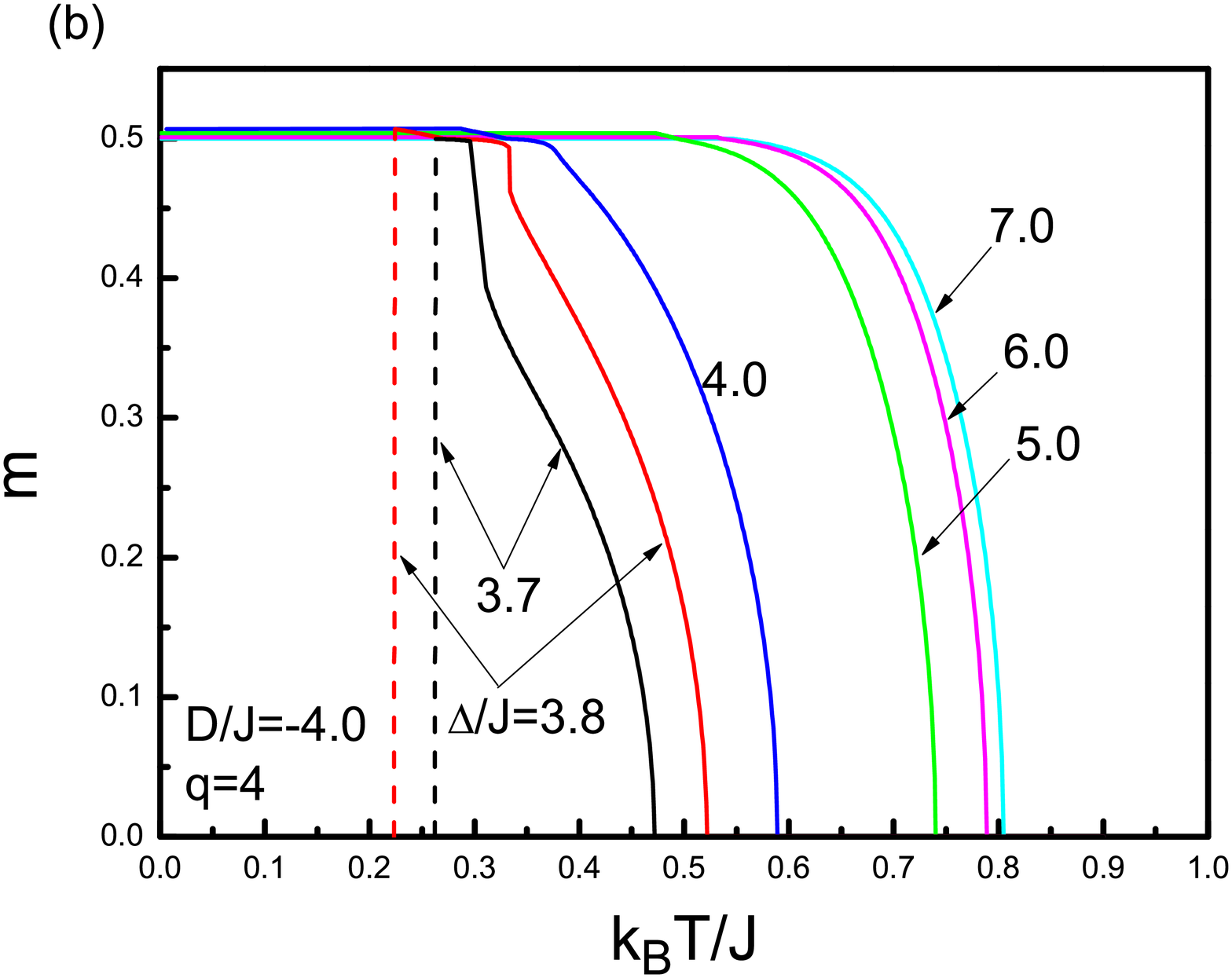}\\
\caption{(Color online) (a) Temperature dependence of magnetization curves for $q=4$ corresponding to Fig. (\ref{fig5}b) for (a) $D/J=2.0$ and (b) $D/J=-4.0$ with some selected values of $\triangle/J$.}\label{fig7}
\end{figure}

\begin{figure}
\includegraphics[width=14cm]{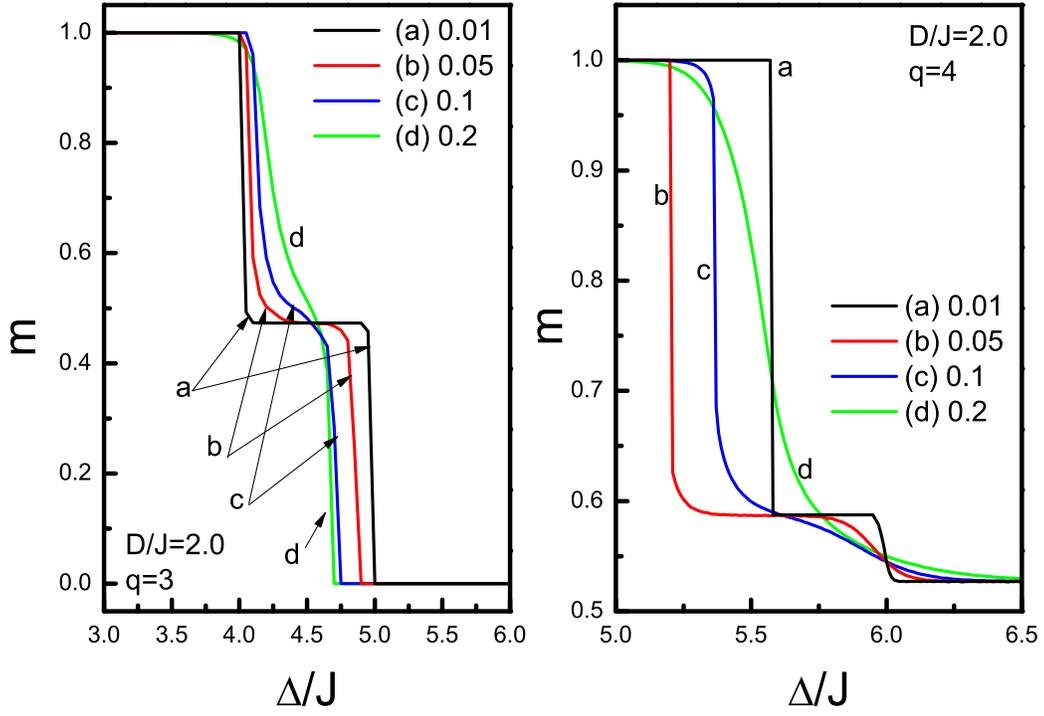}\\
\caption{(Color online) Variation of the magnetization curves as a function of $\triangle/J$ with $D/J=2.0$. Left and right panels are plotted for $q=3$ and $4$, respectively. Each curve is plotted for different temperature values, namely $k_{B}T/J=0.01,0.05,0.1$ and $0.2$.}\label{fig8}
\end{figure}

At the same time, in order to see the effect of the random crystal fields with $\triangle/J,D/J\neq0$ on the phase diagrams and magnetization curves of the system for $q=3$ and $4$, we plot the phase diagrams in $(k_{B}T_{c}/J-D/J)$ plane in Fig. (\ref{fig5}) and variation of the corresponding magnetization curves with temperature in Figs. (\ref{fig6}) and (\ref{fig7}), respectively. At first sight, it is obvious that the phase diagrams in Fig.(\ref{fig5}) represent evident differences in qualitative manner with coordination number $q$. That is, as seen in Fig. (\ref{fig5}a), the curve corresponding to $\triangle/J=0$ represents the phase diagram of pure BC model for a honeycomb lattice which exhibits a reentrant behavior of first order with first and second order transition lines, as well as a tricritical point. From Fig. (\ref{fig5}a), we see that as $\triangle/J$ increases then the tricritical point and first order transitions disappear, and the first order reentrance turns into double reentrance with three transition temperatures of second order, and phase transition lines shift to positive crystal field direction without changing their shapes. On the other hand, the situation is very different for a square lattice. Namely, as seen in Figs. (\ref{fig5}a) and (\ref{fig5}b), $\triangle/J=0$ curves for $q=3$ and $q=4$ are qualitatively identical to each other. However, as seen in Fig. (\ref{fig5}b), for $\triangle/J\neq0$, first order transition lines and tricritical points do not disappear from the system for $q=4$, but shift to negative crystal field values. Besides, the system does not exhibit double reentrance for $q=4$. Furthermore, for $\triangle/J\geq1.5$ in Fig. (\ref{fig5}a), and $\triangle/J\geq0$ in Fig. (\ref{fig5}b), all phase diagrams exhibit similar behavior as $D/J$ varies. Namely, critical temperature $k_{B}T_{c}/J$ in Fig. (\ref{fig5}a) reduces to zero at $D/J=\triangle/J-3.0$. On the other hand, first order transition temperatures in Fig. (\ref{fig5}b), reduce to zero at $D/J=-\triangle/J$.

It is important to note that these observations are consistent with the results shown in Figs. (\ref{fig1}a) and (\ref{fig1}c). In other words, the distribution function given in Eq. (\ref{eq2}) with $p=0.5$ and $D/J=2D_{0}/J$ is identical to Eq. (\ref{eq3}) for $\triangle/J=\pm D_{0}/J$ and $D/J=D_{0}/J$. For example, according to Eq. (\ref{eq2}), if we select $D_{0}/J=4.0$ with $p=0.5$, it means that half of the spins on the lattice sites expose to a crystal field $D/J=0$, while a crystal field given by $D/J=8.0$ acts on the other half of the spins. On the other hand, if we select $\triangle/J=\pm D_{0}/J$ and $D/J=D_{0}/J$ by using Eq. (\ref{eq3}), we generate the same distribution again. Hence, we expect to get the same results in Figs. (\ref{fig1}) and (\ref{fig5}) for these system parameters. For instance, for $D_{0}/J=4.0$ in Fig. (\ref{fig1}a), we get $D/J=8.0$, and the system exhibits a ferromagnetic order in the ground state, which can also be seen in Fig. (\ref{fig5}a) with a critical temperature $k_{B}T_{c}/J=1.4395$. These conditions are also valid for $q=4$, and for the whole temperature region on the phase diagrams. Therefore, the state (para-or ferro), as well as thermal and magnetic properties of a selected $(k_{B}T/J,D/J)$ point with respect to $p=0.5$ curves in Figs. (\ref{fig1}a) and (\ref{fig1}c) is identical to the state of a point $(k_{B}T/J,D/2J)$ in Figs. (\ref{fig5}a) and (\ref{fig5}b) with respect to the curve $\triangle/J=\pm D/J$, respectively. Moreover, the qualitative differences between Figs. (\ref{fig5}a) and (\ref{fig5}b) mentioned above are strongly related to the percolation threshold value of the lattice. Namely, distribution function Eq.(\ref{eq3}) is valid only for $p=0.5$. However, as seen in Table \ref{table1}, we obtain $p_{c}<0.5$ for $q=3$, and $p_{c}>0.5$ for $q=4$.

In Fig. (\ref{fig6}), we examine the temperature dependence of magnetization curves for $q=3$, corresponding to the phase diagrams shown in Fig. (\ref{fig5}a) with $D/J=-1.0$. Fig. (\ref{fig6}a), shows how the temperature dependence of magnetization curves evolve when $\triangle/J$ changes. According to Fig. (\ref{fig6}a), magnetization curves saturate at a partially ordered state at low temperatures. Besides, for $\triangle/J=1.68,1.74$ and $1.80$, the system undergoes three successive phase transitions of second order, which confirms the existence of double reentrance. Similarly, Fig. (\ref{fig6}b) shows how the shape of the magnetization curves change as $D/J$ changes for constant $\triangle/J=6.0$. As seen in Fig. (\ref{fig6}b), magnetization curves exhibit a second order phase transition from a ferromagnetic (fully ordered) to a paramagnetic phase at certain values of crystal field, namely at $D/J=4.0,4.1$ and $4.4$, whereas for $D/J=3.3, 3.5, 3.7$ and $3.9$ the system can only achieve a partially ordered phase. In addition, the curves corresponding to $D/J=3.5, 3.7$ and $3.9$ exhibit a broad maximum at low temperatures, and then decrease as the temperature increases, whereas for $D/J=3.3$, we observe double reentrance. Fig. (\ref{fig7}) shows the magnetization curves for $q=4$, corresponding to the phase diagrams shown in Fig. (\ref{fig5}b). In Fig. (\ref{fig7}a), we see that magnetization curves exhibit a second order phase transition from a paramagnetic phase to a fully ordered ferromagnetic phase for $\triangle/J=0$ and $5.0$. On the other hand, the curves corresponding to $D/J=5.5$, $5.8$ and $6.0$, saturate at a partially ordered state at low temperatures, and exhibit a broad maximum with increasing temperature which depresses gradually as $\triangle/J$ increases, then fall rapidly at a second order phase transition temperature. The broad maximum behavior observed in these curves disappears for $\triangle/J=8.0$. Additionally, Fig. (\ref{fig7}b) represents the magnetization versus temperature curves for $q=4$ with with $D/J=-4.0$. In Fig. (\ref{fig7}b), it is clearly evident that, at low temperatures, the system saturates at a partially ordered phase for $\triangle/J=4.0$, $5.0$, $6.0$ and $7.0$, while for $\triangle/J=3.7$ and $3.8$, a  reentrant behavior of first order occurs. Again we see that,
there is a competition between ferromagnetic exchange interactions and disorder effects in crystal fields which determines the saturation values and temperature dependence of magnetization curves of the system.

Finally, dependence of magnetization of the system on the crystal field $\triangle/J$ for fixed temperature values $k_{B}T/J=0.01,0.05,0.1$ and $0.2$ with $D/J=2.0$ is shown in Fig. (\ref{fig8}) for $q=3$ and $4$, respectively. We see that at sufficiently low temperatures such as $k_{B}T/J=0.01$, magnetization curves exhibit three phases for $q=3$. A first order transition is characterized by a gap in this figure. On the left panel in Fig. (\ref{fig8}), which is plotted for $q=3$, we observe two successive first order phase transitions for $k_{B}T/J=0.01$. The first one is from the fully ordered ferromagnetic phase ($m=1.0$) to the partly ordered phase ($m=0.47$), and the other is from partly ordered phase to disordered phase ($m=0.0$). On the other hand, according to the right panel in Fig. (\ref{fig8}), the system can not reach a paramagnetic phase at the ground state for $q=4$.Hence, we observe two phases. Namely, for $k_{B}T/J=0.01$, a first order phase transition from a fully ordered phase ($m=1.0$) to a partly ordered phase $(m=0.59)$. Then, saturation magnetization of the partly ordered phase reduces continuously to ($m=0.527$). Moreover, the first order transitions disappear with increasing temperatures, both for $q=3$ and $4$. Since in Figs. (\ref{fig5}a) and (\ref{fig5}b), all phase diagrams exhibit similar behavior as $D/J$ varies, behavior of magnetization curves in Fig. (\ref{fig8}) should be the same for different $D/J$ values. Namely, the left panel of Fig. (\ref{fig8}) can be regarded as the variation of magnetization with $\triangle/J$ within the range  $D/J+1.0<\triangle/J<D/J+4.0$ for $q=3$. for a selected value of $D/J$. It is possible to observe the similar behavior on the right panel in Fig. (\ref{fig8}). This general behavior is an expected result, since the behavior of the system depends on the relation between $\triangle/J$ and $D/J$ parameters for a given temperature, not their values independently.

\section{Conclusions}\label{conclude}
In this work, we have studied the phase diagrams of a spin-1 Blume-Capel model with diluted and random crystal field interactions on two dimensional lattices. We have introduced an effective-field approximation that takes into account the correlations between different spins in the cluster of a considered lattice and examined the phase diagrams as well as magnetization curves of the system for different types of crystal field distributions, namely, dilute crystal fields and a double peaked delta  distribution, given by Eqs. (\ref{eq2}) and (\ref{eq3}), respectively.

For dilute crystal fields, we have given a detailed exploration of the global phase diagrams of the system in $k_{B}T_{c}/J-D/J$ plane with the second and first order transitions, as well as tricritical points. We have also shown that the system with dilute crystal fields exhibits a percolation threshold value $p_{c}$ which can not be predicted by standard MFA. In addition, we have observed multi-reentrant phase transitions for specific set of system parameters.

On the other hand, we have investigated the effect of the random crystal field distribution characterized by two crystal field parameters $D/J$ and $\triangle/J$ on the phase diagrams of the system. As a limited case, we have also focused on a bimodal distribution with $D/J=0$. Particulary, we have reported the following observations for a bimodal distribution: It has been found that the phase diagrams have symmetric shape with respect to $\triangle/J$ which comes from the fact that $p=1/2$. The transition temperatures are of second order, and the system exhibit different characteristic features depending on the coordination number $q$. Besides, we have realized that the system may exhibit clear distinctions in qualitative manner with coordination number $q$ for random crystal fields with $\triangle/J,D/J\neq0$. Moreover, we have discussed a competition effect which arises from the presence of dilution, as well as random crystal fields, and we have observed that saturation values of the magnetization curves are strongly related to these effects.

As a result, we can conclude that all of the points mentioned above show that our method improves the conventional EFT methods based on decoupling approximation. Therefore, we hope that the results obtained in this work may be beneficial from both theoretical and experimental points of view.

\section*{Acknowledgements}
One of the authors (Y.Y.) would like to thank the Scientific and Technological Research Council of Turkey (T\"{U}B\.{I}TAK) for partial financial support. This work has been completed at Dokuz Eyl\"{u}l University, Graduate School of Natural and Applied Sciences. Partial financial support from SRF (Scientific Research Fund) of Dokuz Eyl\"{u}l University (2009.KB.FEN.077) (H.P.) is also acknowledged.

\newpage
\appendix
\section{Fundamental correlation functions of the system for a square lattice}\label{appendixa}
Magnetization of the central spin for a square lattice is given as follows
\begin{eqnarray}\label{eqA1}
\nonumber
\left\langle\langle S_{0}^{z}\right\rangle\rangle&=&\mu_{0}+4c_{1}\langle\langle S_{1}\rangle\rangle_{r}+4(\mu_{2}-\mu_{0})\langle\langle S_{1}^{2}\rangle\rangle_{r}\\
\nonumber
&&+6\mu_{1}\langle\langle S_{1}S_{2}\rangle\rangle_{r}+12(c_{2}-c_{1})\langle\langle S_{1}S_{2}^{2}\rangle\rangle_{r}\\
\nonumber
&&+6(\mu_{0}-2\mu_{2}+\mu_{3})\langle\langle S_{1}^{2}S_{2}^{2}\rangle\rangle_{r}+4c_{3}\langle\langle S_{1}S_{2}S_{3}\rangle\rangle_{r}\\
\nonumber
&&+12(\mu_{4}-\mu_{1})\langle\langle S_{1}S_{2}S_{3}^{2}\rangle\rangle_{r}+12(c_{4}-2c_{2}+c_{1})\langle\langle S_{1}S_{2}^{2}S_{3}^{2}\rangle\rangle_{r}\\
\nonumber
&&+4(\mu_{5}-3\mu_{3}+3\mu_{2}-\mu_{0})\langle\langle S_{1}^{2}S_{2}^{2}S_{3}^{2}\rangle\rangle_{r}+\mu_{8}\langle\langle S_{1}S_{2}S_{3}S_{4}\rangle\rangle_{r}\\
\nonumber
&&+4(c_{5}-c_{3})\langle\langle S_{1}S_{2}S_{3}S_{4}^{2}\rangle\rangle_{r}+6(\mu_{1}-2\mu_{4}+\mu_{6})\langle\langle S_{1}S_{2}S_{3}^{2}S_{4}^{2}\rangle\rangle_{r}\\
\nonumber
&&+4(c_{6}-3c_{4}+3c_{2}-c_{1})\langle\langle S_{1}S_{2}^{2}S_{3}^{2}S_{4}^{2}\rangle\rangle_{r}\\
&&+(\mu_{0}-4\mu_{2}+6\mu_{3}-4\mu_{5}+\mu_{7})\langle\langle S_{1}^{2}S_{2}^{2}S_{3}^{2}S_{4}^{2}\rangle\rangle_{r},
\end{eqnarray}
where the coefficients are given by
\begin{eqnarray}
\nonumber
\mu_{0}=F(0),\ \ \ \ \ \ \ \ \ \ \ \ \ \ \ \ \ \ \ \ \ \ \ \ \ \ \ \ \ \ \ \ &&\\
\nonumber
\mu_{1}=\sinh^{2}(J\nabla)F(x)|_{x=0},\ \ \ \ \ \ \ \ \ \ \ \ \ &&c_{1}=\sinh(J\nabla)F(x)_{x=0},\\
\nonumber
\mu_{2}=\cosh(J\nabla)F(x)|_{x=0},\ \ \ \ \ \ \ \ \ \ \ \ \ \ &&c_{2}=\sinh(J\nabla)\cosh(J\nabla)F(x)_{x=0},\\
\nonumber
\mu_{3}=\cosh^{2}(J\nabla)F(x)|_{x=0},\ \ \ \ \ \ \ \ \ \ \ \ \ &&c_{3}=\sinh^{3}(J\nabla)F(x)_{x=0},\\
\nonumber
\mu_{4}=\sinh^{2}(J\nabla)\cosh(J\nabla)F(x)|_{x=0},&&c_{4}=\cosh^{2}(J\nabla)\sinh(J\nabla)F(x)_{x=0},\\
\nonumber
\mu_{5}=\cosh^{3}(J\nabla)F(x)|_{x=0},\ \ \ \ \ \ \ \ \ \ \ \ \ &&c_{5}=\sinh^{3}(J\nabla)\cosh(J\nabla)F(x)_{x=0},\\
\nonumber
\mu_{6}=\sinh^{2}(J\nabla)\cosh^{2}(J\nabla)F(x)_{x=0},&&c_{6}=\cosh^{3}(J\nabla)\sinh(J\nabla)F(x)_{x=0},\\
\nonumber
\mu_{7}=\cosh^{4}(J\nabla)F(x)|_{x=0},\ \ \ \ \ \ \ \ \ \ \ \ \ &&\\
\nonumber
\mu_{8}=\sinh^{4}(J\nabla)F(x)_{x=0}.\ \ \ \ \ \ \ \ \ \ \ \ \ \
\end{eqnarray}
Quadrupolar moment corresponding to equation (\ref{eqB1}) defined as
\begin{eqnarray}\label{eqA2}
\nonumber
\left\langle\langle (S_{0}^{z})^{2}\right\rangle\rangle&=&\rho_{0}+4\eta_{1}\langle\langle S_{1}\rangle\rangle_{r}+4(\rho_{2}-\rho_{0})\langle\langle S_{1}^{2}\rangle\rangle_{r}\\
\nonumber
&&+6\rho_{1}\langle\langle S_{1}S_{2}\rangle\rangle_{r}+12(\eta_{2}-\eta_{1})\langle\langle S_{1}S_{2}^{2}\rangle\rangle_{r}\\
\nonumber
&&+6(\rho_{0}-2\rho_{2}+\rho_{3})\langle\langle S_{1}^{2}S_{2}^{2}\rangle\rangle_{r}+4\eta_{3}\langle\langle S_{1}S_{2}S_{3}\rangle\rangle_{r}\\
\nonumber
&&+12(\rho_{4}-\rho_{1})\langle\langle S_{1}S_{2}S_{3}^{2}\rangle\rangle_{r}+12(\eta_{4}-2\eta_{2}+\eta_{1})\langle\langle S_{1}S_{2}^{2}S_{3}^{2}\rangle\rangle_{r}\\
\nonumber
&&+4(\rho_{5}-3\rho_{3}+3\rho_{2}-\rho_{0})\langle\langle S_{1}^{2}S_{2}^{2}S_{3}^{2}\rangle\rangle_{r}+\rho_{8}\langle\langle S_{1}S_{2}S_{3}S_{4}\rangle\rangle_{r}\\
\nonumber
&&+4(\eta_{5}-\eta_{3})\langle\langle S_{1}S_{2}S_{3}S_{4}^{2}\rangle\rangle_{r}+6(\rho_{1}-2\rho_{4}+\rho_{6})\langle\langle S_{1}S_{2}S_{3}^{2}S_{4}^{2}\rangle\rangle_{r}\\
\nonumber
&&+4(\eta_{6}-3\eta_{4}+3\eta_{2}-\eta_{1})\langle\langle S_{1}S_{2}^{2}S_{3}^{2}S_{4}^{2}\rangle\rangle_{r}\\
&&+(\rho_{0}-4\rho_{2}+6\rho_{3}-4\rho_{5}+\rho_{7})\langle\langle S_{1}^{2}S_{2}^{2}S_{3}^{2}S_{4}^{2}\rangle\rangle_{r},
\end{eqnarray}
where
\begin{eqnarray}
\nonumber
\rho_{0}=G(0),\ \ \ \ \ \ \ \ \ \ \ \ \ \ \ \ \ \ \ \ \ \ \ \ \ \ \ \ \ \ \ \ &&\\
\nonumber
\rho_{1}=\sinh^{2}(J\nabla)G(x)|_{x=0},\ \ \ \ \ \ \ \ \ \ \ \ \ &&\eta_{1}=\sinh(J\nabla)G(x)_{x=0},\\
\nonumber
\rho_{2}=\cosh(J\nabla)G(x)|_{x=0},\ \ \ \ \ \ \ \ \ \ \ \ \ \ &&\eta_{2}=\sinh(J\nabla)\cosh(J\nabla)G(x)_{x=0},\\
\nonumber
\rho_{3}=\cosh^{2}(J\nabla)G(x)|_{x=0},\ \ \ \ \ \ \ \ \ \ \ \ \ &&\eta_{3}=\sinh^{3}(J\nabla)G(x)_{x=0},\\
\nonumber
\rho_{4}=\sinh^{2}(J\nabla)\cosh(J\nabla)G(x)|_{x=0},&&\eta_{4}=\cosh^{2}(J\nabla)\sinh(J\nabla)G(x)_{x=0},\\
\nonumber
\rho_{5}=\cosh^{3}(J\nabla)G(x)|_{x=0},\ \ \ \ \ \ \ \ \ \ \ \ \ &&\eta_{5}=\sinh^{3}(J\nabla)\cosh(J\nabla)G(x)_{x=0},\\
\nonumber
\rho_{6}=\sinh^{2}(J\nabla)\cosh^{2}(J\nabla)G(x)_{x=0},&&\eta_{6}=\cosh^{3}(J\nabla)\sinh(J\nabla)G(x)_{x=0},\\
\nonumber
\rho_{7}=\cosh^{4}(J\nabla)G(x)|_{x=0},\ \ \ \ \ \ \ \ \ \ \ \ \ &&\\
\nonumber
\rho_{8}=\sinh^{4}(J\nabla)G(x)_{x=0}.\ \ \ \ \ \ \ \ \ \ \ \ \ \
\end{eqnarray}
Finally, perimeter spin identities are as follows
\begin{eqnarray}\label{eqA3}
\langle\langle S_{1}\rangle\rangle_{r}&=&\alpha_{1}(1-\langle\langle (S_{0})^{2}\rangle\rangle_{r})+\alpha_{2}\langle\langle S_{0}\rangle\rangle_{r}+\alpha_{3}\langle\langle (S_{0})^{2}\rangle\rangle_{r},\\
\langle\langle S_{1}^{2}\rangle\rangle_{r}&=&\omega_{1}+\omega_{2}\langle\langle S_{0}\rangle\rangle_{r}+(\omega_{3}-\omega_{1})\langle\langle S_{0}^{2}\rangle\rangle_{r}
\end{eqnarray}
with the coefficients
\begin{eqnarray}
\nonumber
\alpha_{1}=F(\gamma)\ \ \ \ \ \ \ \ \ \ \ \ \ \ \ \ \ \ &&\omega_{1}=G(\gamma)|_{x=0}\\
\nonumber
\alpha_{2}=\sinh(J\nabla)F(x+\gamma)&&\omega_{2}=\sinh(J\nabla)G(x+\gamma)|_{x=0}\\
\nonumber
\alpha_{3}=\cosh(J\nabla)F(x+\gamma)&&\omega_{3}=\cosh(J\nabla)G(x+\gamma)|_{x=0}
\end{eqnarray}
where $\gamma=(q-1)A$ with $q=4$, and the functions $F(x)$ and $G(x)$ are defined in equations (\ref{eq9}) and (\ref{eq16}).
\section{The complete set of twenty one linear equations of a honeycomb lattice
}\label{appendixb}
\begin{eqnarray}\label{eqB1}
\nonumber\langle\langle
S_{0}^{z}\rangle\rangle_{r}&=&l_{0}+3k_{1}\langle\langle S_{1}
\rangle\rangle_{r}+3(l_{1}-l_{0})\langle\langle S_{1}^{2}
\rangle\rangle_{r}+3l_{2}\langle\langle S_{1}S_{2}
\rangle\rangle_{r} \\
\nonumber& & +6(k_{2}-k_{1})\langle\langle S_{1}S_{2}^{2}
\rangle\rangle_{r}+3(l_{0}-2l_{1}+l_{3})\langle\langle
S_{1}^{2}S_{2}^{2}
\rangle\rangle_{r} \\
\nonumber& &+k_{3}\langle\langle S_{1}S_{2}S_{3}
\rangle\rangle_{r}+3(l_{4}-l_{2})\langle\langle S_{1}S_{2}S_{3}^{2}
\rangle\rangle_{r}\\
\nonumber& &+3(k_{1}-2k_{2}+k_{4})\langle\langle
S_{1}S_{2}^{2}S_{3}^{2}
\rangle\rangle_{r}\\
\nonumber& &+(-l_{0}+3l_{1}-3l_{3}+l_{5})\langle\langle
S_{1}^{2}S_{2}^{2}S_{3}^{2} \rangle\rangle_{r}\\
\nonumber \langle\langle S_{1}S_{0}\rangle\rangle_{r} &=&
(3l_{1}-2l_{0})\langle\langle
S_{1}\rangle\rangle_{r}+3k_{1}\langle\langle
S_{1}^{2}\rangle\rangle_{r}+3(l_{0}-2l_{1}+l_{2}+l_{3})\langle\langle
S_{1}S_{2}^{2}\rangle\rangle_{r}  \\
\nonumber&&+6(k_{2}-k_{1})\langle\langle
S_{1}^2S_{2}^{2}\rangle\rangle_{r}+k_{3}\langle\langle
S_{1}S_{2}S_{3}^{2}\rangle\rangle_{r}\\
\nonumber&&+(-l_{0}+3l_{1}-3l_{2}-3l_{3}+3l_{4}+l_{5})\langle\langle
S_{1}S_{2}^{2}S_{3}^{2}\rangle\rangle_{r}\\
\nonumber&&+3(k_{1}-2k_{2}+k_{4})\langle\langle
S_{1}^{2}S_{2}^{2}S_{3}^{2}\rangle\rangle_{r}\\
\nonumber \langle\langle
S_{1}S_{2}S_{0}\rangle\rangle_{r} &=& ({l_0}-3{l_1}+3{l_2}+3{l_3})\langle\langle{S_1}{S_2}\rangle\rangle_{r}+(6{k_2}-3{k_1})\langle\langle{S_1}S_{2}^{2}\rangle\rangle_{r} \\
\nonumber&&+
(-{l_0}+3{l_1}-3{l_2}-3{l_3}+3{l_4}+{l_5})\langle\langle{S_1}{S_2}S_{3}^{2}\rangle\rangle_{r}\\
\nonumber&&+(3{k_1}-6{k_2}+{k_3}+3{k_4})\langle\langle{S_1}S_{2}^{2}S_{3}^{2}\rangle\rangle_{r}\\
\nonumber \langle\langle S_{1}\rangle\rangle_{r} &=&
a_{1}(1-\langle\langle
(S_{0}^{z})^{2}\rangle\rangle_{r})+a_{2}\langle\langle S_{0}^{z}
\rangle\rangle_{r}+a_{3}\langle\langle
(S_{0}^{z})^{2}\rangle\rangle_{r} \\
\nonumber  \langle\langle S_{1}S_{2}\rangle\rangle_{r} &=& a_{1}
\langle\langle S_{1}
\rangle\rangle_{r}+a_{2}\langle\langle S_{0}S_{1}\rangle\rangle_{r}+(a_{3}-a_{1})\langle\langle S_{1}S_{0}^{2}\rangle\rangle_{r} \\
\nonumber  \langle\langle S_{1}S_{2}S_{3}\rangle\rangle_{r} &=&
a_{1}\langle\langle S_{1}S_{2}\rangle\rangle_{r}
+a_{2}\langle\langle S_{0}S_{1}S_{2}\rangle\rangle_{r}+(a_{3}-a_{1})\langle\langle S_{1}S_{2}S_{0}^{2}\rangle\rangle_{r}  \\
\nonumber \langle\langle
S_{1}^{2}\rangle\rangle_{r}&=&b_{1}(1-\langle\langle
(S_{0}^{z})^{2}\rangle\rangle_{r})+b_{2}\langle\langle S_{0}^{z}
\rangle\rangle_{r}+b_{3}\langle\langle
(S_{0}^{z})^{2}\rangle\rangle_{r}\\
\nonumber  \langle\langle S_{1}S_{2}^2\rangle\rangle_{r} &=&
b_{1}\langle\langle S_{1}
\rangle\rangle_{r}+b_{2}\langle\langle S_{0}S_{1}\rangle\rangle_{r}+(b_{3}-b_{1}) \langle\langle S_{1}S_{0}^2\rangle\rangle_{r}  \\
\nonumber  \langle\langle S_{1}^2S_{2}^2\rangle\rangle_{r} &=&
b_{1}\langle\langle S_{1}^2\rangle\rangle_{r}+b_{2}
\langle\langle S_{0}S_{1}^2\rangle\rangle_{r}+(b_{3}-b_{1})\langle\langle S_{1}^2S_{0}^2\rangle\rangle_{r}  \\
\nonumber  \langle\langle S_{0}S_{1}^2\rangle\rangle_{r} &=& b_{3}\langle\langle S_{0}\rangle\rangle_{r}+b_{2}\langle\langle S_{0}^2\rangle\rangle_{r}  \\
\nonumber  \langle\langle S_{0}S_{1}S_{2}^2\rangle\rangle_{r} &=& b_{3}\langle\langle S_{0}S_{1}\rangle\rangle_{r}+b_{2}\langle\langle S_{1}S_{0}^2\rangle\rangle_{r} \\
\nonumber  \langle\langle S_{0}S_{1}^2S_{2}^2\rangle\rangle_{r} &=& b_{3}\langle\langle S_{0}S_{1}^2\rangle\rangle_{r}+b_{2}\langle\langle S_{1}^2S_{0}^2\rangle\rangle_{r} \\
\nonumber  \langle\langle S_{1}S_{2}S_{3}^2\rangle\rangle_{r} &=&
b_{1}\langle\langle S_{1}S_{2}\rangle\rangle_{r}
+b_{2}\langle\langle S_{0}S_{1}S_{2}\rangle\rangle_{r}+(b_{3}-b_{1})\langle\langle S_{1}S_{2}S_{0}^2\rangle\rangle_{r}  \\
\nonumber  \langle\langle S_{1}S_{2}^2S_{3}^2\rangle\rangle_{r} &=&
b_{1}\langle\langle S_{1}S_{2}^2\rangle\rangle_{r}
+b_{2}\langle\langle S_{0}S_{1}S_{2}^2\rangle\rangle_{r}+(b_{3}-b_{1})\langle\langle S_{1}S_{2}^2S_{0}^2\rangle\rangle_{r} \\
\nonumber  \langle\langle S_{1}^2S_{2}^2S_{3}^2\rangle\rangle_{r}
&=& b_{1}\langle\langle S_{1}^2S_{2}^2\rangle\rangle_{r}
+b_{2}\langle\langle S_{0}S_{1}^2S_{2}^2\rangle\rangle_{r}+(b_{3}-b_{1})\langle\langle S_{1}^2S_{2}^2S_{0}^2\rangle\rangle_{r} \\
\nonumber  \langle\langle S_{0}^2\rangle\rangle_{r} &=&
r_{0}+3n_{1}\langle\langle S_{1}\rangle\rangle_{r}
+3(r_{1}-r_{0})\langle\langle S_{1}^{2}\rangle\rangle_{r}+3r_{2}\langle\langle S_{1}S_{2}\rangle\rangle_{r}  \\
\nonumber   &&+6(n_{2}-n_{1})\langle\langle
S_{1}S_{2}^{2}\rangle\rangle_{r}+3(r_{0}-2r_{1}+r_{3})\langle\langle
S_{1}^{2}S_{2}^{2}\rangle\rangle_{r}
+n_{3}\langle\langle S_{1}S_{2}S_{3}\rangle\rangle_{r}  \\
\nonumber   &&+ 3(r_{4}-r_{2})\langle\langle
S_{1}S_{2}S_{3}^{2}\rangle\rangle_{r}
+3(n_{1}-2n_{2}+n_{4})\langle\langle S_{1}S_{2}^{2}S_{3}^{2}\rangle\rangle_{r}  \\
\nonumber&&+(-r_{0}+3r_{1}-3r_{3}+r_{5})\langle\langle
S_{1}^{2}S_{2}^{2}S_{3}^{2}\rangle\rangle_{r}\\
\nonumber  \langle\langle S_{1}S_{0}^2\rangle\rangle_{r}
&=&(3r_{1}-2r_{0})\langle\langle S_{1}\rangle\rangle_{r}
+3n_{1}\langle\langle S_{1}^2\rangle\rangle_{r}+(3r_{2}+3r_{0}-6r_{1}+3r_{3})\langle\langle S_{1}S_{2}^2\rangle\rangle_{r}  \\
&&\nonumber +6(n_{2}-n_{1})\langle\langle
S_{1}^2S_{2}^2\rangle\rangle_{r}+n_{3}\langle\langle
S_{1}S_{2}S_{3}^2\rangle\rangle_{r}\\
&&\nonumber+(-r_{0}+3r_{1}-3r_{2}-3r_{3}+3r_{4}+r_{5})\langle\langle
S_{1}S_{2}^2S_{3}^2\rangle\rangle_{r}\\
&&\nonumber+3(n_{1}-2n_{2}+n_{4})\langle\langle S_{1}^2S_{2}^2S_{3}^2\rangle\rangle_{r}\\
\nonumber  \langle\langle S_{1}^2S_{0}^2\rangle\rangle_{r} &=&
(3r_{1}-2r_{0})\langle\langle S_{1}^2\rangle\rangle_{r}
+3n_{1}\langle\langle S_{1}\rangle\rangle_{r}+(3r_{2}+3r_{0}-6r_{1}+3r_{3})\langle\langle S_{1}^2S_{2}^2\rangle\rangle_{r} \\
\nonumber&&+6(n_{2}-n_{1})\langle\langle S_{1}S_{2}^2\rangle\rangle_{r}+(3n_{1}-6n_{2}+n_{3}+3n_{4})\langle\langle S_{1}S_{2}^2S_{3}^2\rangle\rangle_{r}\\
\nonumber&&+(-r_{0}+3r_{1}-3r_{2}-3r_{3}+3r_{4}+r_{5})\langle\langle
S_{1}^2S_{2}^2S_{3}^2\rangle\rangle_{r}\\
\nonumber  \langle\langle S_{1}S_{2}S_{0}^2\rangle\rangle_{r} &=&
(r_{0}-3r_{1}
+3r_{2}+3r_{3})\langle\langle S_{1}S_{2}\rangle\rangle_{r}+(-3n_{1}+6n_{2})\langle\langle S_{1}S_{2}^2\rangle\rangle_{r} \\
\nonumber&& (-r_{0}+3r_{1}-3r_{2}-3r_{3}+3r_{4}+r_{5})\langle\langle
S_{1}S_{2}S_{3}^2\rangle\rangle_{r} \\
\nonumber&&+(3n_{1}-6n_{2}+n_{3}+3n_{4})\langle\langle S_{1}S_{2}^2S_{3}^2\rangle\rangle_{r}\\
\nonumber  \langle\langle S_{1}S_{2}^2S_{0}^2\rangle\rangle_{r}
&=&(r_{0}-3r_{1}
+3r_{2}+3r_{3})\langle\langle S_{1}S_{2}^2\rangle\rangle_{r}+(-3n_{1}+6n_{2})\langle\langle S_{1}S_{2}\rangle\rangle_{r} \\
\nonumber&& (-r_{0}+3r_{1}-3r_{2}-3r_{3}+3r_{4}+r_{5})\langle\langle
S_{1}S_{2}^2S_{3}^2\rangle\rangle_{r} \\
\nonumber&&+(3n_{1}-6n_{2}+n_{3}+3n_{4})\langle\langle S_{1}S_{2}S_{3}^2\rangle\rangle_{r}\\
\nonumber \langle\langle
S_{1}^{2}S_{2}^{2}S_{0}^{2}\rangle\rangle_{r}
\nonumber&=&({r_0}-3{r_1}+3{r_2}+3{r_3})\langle\langle
S_{1}^{2}S_{2}^{2}\rangle\rangle_{r}+(-3{n_1}+6{n_2})\langle\langle
{S_1}S_{2}^{2}\rangle\rangle_{r}\\
\nonumber&&+(3{n_1}-6{n_2}+{n_3}+3{n_4})\langle\langle
{S_1}S_{2}^{2}S_{3}^{2}\rangle\rangle_{r}\\
\nonumber&&+(-{r_0}+3{r_1}-3{r_2}-3{r_3}+3{r_4}+{r_5})\langle\langle
S_{1}^{2}S_{2}^{2}S_{3}^{2}\rangle\rangle_{r}.\\
\end{eqnarray}

\newpage

\newpage
\end{document}